\documentstyle[prd,aps,eqsecnum,preprint,tighten,psfig,floats]{revtex}
\begin{document}
\draft

\preprint{ITP-SB-94-39}

\title{Anomalous dimension of the gluon operator \\ in pure Yang-Mills theory}

\author{B. W. Harris and J. Smith}

\address{Institute for Theoretical Physics, State University of 
New York at Stony Brook, \\ Stony Brook, NY 11794-3840, USA}

\date{\today}

\maketitle

\begin{abstract}
We present new one loop calculations that confirm the theorems 
of Joglekar and Lee on the renormalization of composite operators.
We do this by considering physical matrix elements 
with the operators inserted at {\em non-zero} momentum.  The
resulting IR singularities are regulated dimensionally.  
We show that the physical matrix element of the BRST exact gauge 
variant operator which appears in the energy-momentum tensor is zero.
We then show that the physical matrix elements of the classical 
energy-momentum tensor and the gauge invariant twist two gluon operator are 
independent of the gauge fixing parameter.  A Sudakov factor 
appears in the latter cases.  The universality of 
this factor and the UV finiteness of the energy-momentum tensor 
provide another method of finding the anomalous dimension of the 
gluon operator. We conjecture that this method applies to higher loops and 
takes full advantage of the triangularity of the mixing matrix.
\end{abstract}


\section{Introduction}
The anomalous dimension of the twist two 
gluon operator \cite{gw} is an important 
quantity in QCD phenomenology and has been the subject of study for 
many years.  It contributes to the logarithmic violation of Bjorken 
scaling which was one of the great early successes of QCD.  Its 
inverse Mellin 
transform is the Gribov-Lipatov-Altarelli-Parisi gluon splitting kernel.
In the late 1970's two groups independently calculated 
the anomalous dimension of the gluon operator in second order. 
One group worked in the axial gauge \cite{fur} and the other in the 
Feynman gauge \cite{sac}.  The results were 
different, and the questionable axial gauge result was 
chosen to be correct on the basis of 
a relation among anomalous dimensions derived from supersymmetry \cite{susy}.
The third order anomalous dimension of the gluon operator 
is presently needed to complete the second order QCD corrections to the 
deep inelastic structure functions $F_2(x,Q^2)$ and $F_L(x,Q^2)$ \cite{ZvN}. 
In preparation for this third order calculation one must completely understand 
the discrepancies between results of the  two loop calculations. 
Therefore Hamberg and van Neerven \cite{ham} recently recalculated 
the anomalous dimension of the gluon operator to second order in 
the Feynman gauge. Their result 
agreed with the previous axial gauge result.
However, the situation is still not entirely satisfactory,  
as it is not obvious that the new calculation satisfies
the general theorems on the renormalization 
of composite operators in gauge theories \cite{jog}, \cite{clee}.

There are three points of contention.  
First, the physical matrix elements of the gauge variant (GV) 
operators \cite{dt} used by Hamberg and van Neerven
do not vanish.
Second, a physical matrix element of the gauge invariant (GI) 
gluon operator turns out to be gauge dependent.  
Third, their renormalization mixing matrix is not triangular.
While these findings sound bad, they do not contradict the general theorems 
\cite{jog}, \cite{clee} which state that the operators should be 
inserted at non-zero momentum.  
Since Hamberg and van Neerven worked with operator insertions at 
zero momentum the general theorems do not apply.  This 
point has recently been emphasized by Collins and Scalise \cite{randy} who 
isolated an infra-red (IR) singularity in the proof that the physical 
matrix elements of BRST exact operators vanish.

To clarify the situation
we consider it necessary to verify the general theorems by explicit 
calculations.  To our knowledge this has never been done mostly because 
there was no clear way to handle the IR singularities encountered when 
putting the external legs on-mass-shell \cite{jog}, \cite{clee}.  The 
on-mass-shell renormalization scheme is not well defined in QCD but one 
may try a variant of this scheme wherein the external legs are put 
on-mass-shell in the limiting sense only.  This leads to the so called 
``modified LSZ'' prescription \cite{randy}.  
We show below that if one tries to verify the gauge parameter 
independence of a physical matrix element 
of the gauge invariant classical energy-momentum tensor
inserted at non-zero momentum using this method the gauge dependence remains 
and the result contains logarithms of the external momentum squared.  
Hence the IR singularities, which
arise when putting the external momenta on-mass-shell, 
spoil the gauge independence of the result.  We present
a calculation in which we resolve this problem.
We have found, by inserting the GI operators at 
non-zero momentum and using dimensional regularization to regulate the 
IR divergences, that there are universal double 
and single IR pole structures in the physical matrix elements of both the 
energy momentum tensor and the gluon operator.  
Performing the calculation in this way in space-time dimension $n>4$, 
we verify the general theorems.
By appealing to the ultra-violet (UV)
finiteness of the energy-momentum tensor, we find that the 
difference in the coefficients of the single pole terms is 
precisely the anomalous dimension of the twist two gluon operator.
This approach is very similar to the factorization of 
mass singularities in perturbative QCD and the way one deals with Sudakov
logarithms.  A review of such logarithmic factors is given
in \cite{sudakov}.  We believe this method will also work 
for the anomalous dimension in higher loops and an all-orders
proof of the IR factorization should be possible.

This paper is organized as follows.  In Section II we introduce
our notation for the operators.  
The Feynman rules are given in Appendix A.  In Section III we 
list some of the predictions of the general theorems on the 
renormalization of composite operators.  In the main section, Section IV, we 
present an explicit first order calculation of the physical matrix 
elements with the energy-momentum tensor and the twist two 
gluon operator inserted at non-zero momentum using both of the 
methods introduced above.
Appendix B contains information on the evaluation of the integrals 
needed to perform the calculations.  We discuss our findings and give 
our conclusions in Section V.

\section{Notation}
We consider the effective Lagrangian
\begin{equation}
{\cal L} = {\cal L}_{YM} + {\cal L}_{gf} + {\cal L}_{FP} \,,
\end{equation}
where ${\cal L}_{YM}$, ${\cal L}_{gf}$, and ${\cal L}_{FP}$ are the Yang-Mills, gauge 
fixing, and Faddeev-Popov ghost Lagrangians, respectively, and are 
given by
\begin{eqnarray}
{\cal L}_{YM} & = & 
	- \frac{1}{4} F_{\mu \nu}^{a} F^{a \mu \nu} \, , \\ \nonumber
{\cal L}_{gf} & = & 
	- \frac{1}{2 \alpha} ( \partial^{\mu} A^{a}_{\mu} )^2 \, , \\ \nonumber
{\cal L}_{FP} & = & ( \partial^{\mu} \xi^{a} ) D^{ab}_{\mu} \omega^{b} \,.
\end{eqnarray}
In this paper we denote renormalized quantities by the subscript $r$.  For 
example, $A^a_{\mu,r}$ is the renormalized Yang-Mills (YM) field and 
$A^a_\mu $ is the bare YM field, $\omega^a$ is the bare anti-commuting ghost 
field, and $\xi^a$ is the bare anti-commuting anti-ghost field.  
The non-Abelian 
field strength $F^a_{\mu \nu}$ and the covariant derivative $D^{ab}_\mu$ 
are defined in terms of bare quantities as
\begin{eqnarray}
F^{a}_{\mu \nu} & = & \partial_{\mu} A^{a}_{\nu} - \partial_{\nu} A^{a}_{\mu}
			+ g f^{abc} A^{b}_{\mu} A^{c}_{\nu} \, , \\ \nonumber
D_{\mu}^{ab} 	& = & \delta^{ab} \partial_{\mu} - g f^{abc} A^{c}_{\mu}\,.
\end{eqnarray}
The structure constants of $SU(3)$ are defined by 
$[ T^a, T^b ] = i f^{abc} T^c$ where $T^a$ are the
generators of $SU(3)$.  The bare gauge 
fixing parameter is $\alpha$, and the bare coupling constant is $g$.
The YM Lagrangian is (constructed to be) invariant under the 
infinitesimal gauge transformation
\begin{equation}
\delta A_{\mu}^{a} = - \frac{1}{g} D_{\mu}^{ab} \theta^{b}\,,
\end{equation}
while the effective Lagrangian (2.1) is invariant under the 
infinitesimal BRST \cite{brs}, \cite{t} transformation
\begin{equation}
\delta_{BRST} A_{\mu}^{a} = \lambda D^{ab}_{\mu} \omega^{b}\,,
\end{equation}
with $\lambda$ anti-commuting, provided the ghost and 
anti-ghost transform as
\begin{eqnarray}
\delta_{BRST} \omega^{a} & = & - \frac{1}{2} g f^{abc}\omega^{b}\omega^{c}
 \lambda \, , \\ 
\nonumber
\delta_{BRST} \xi^{a}	  & = & - \frac{1}{\alpha} ( \partial^{\mu} A^a_{\mu} ) 
\lambda \,.
\end{eqnarray}

The energy-momentum tensor \cite{clee} \cite{fmw} follows from the effective 
Lagrangian by differentiation with respect to the metric tensor \cite{LL} 
without use of the equations of motion.  
It is then contracted
with two powers of an arbitrary, light-like vector $\Delta^{\mu}$ to 
pick out the symmetric traceless part.  We define
\begin{equation}
O_{em}  \equiv  - \frac{1}{2} \Delta^{\mu} \Delta^{\nu} \theta_{\mu \nu} = 
O_{em}^{GI} + O_{em}^{GV} \,,
\end{equation}
where
\begin{eqnarray}
O_{em}^{GI} & = & - \frac{1}{2} \Delta^{\mu} \Delta^{\nu} 
	\theta_{\mu \nu}^{GI} = \frac{1}{2} \Delta^{\mu} 
	F_{\mu \rho}^{a} \Delta^{\nu} F^{a \; \rho}_{\nu} \, , \\
O_{em}^{GV} & = & - \frac{1}{2} \Delta^{\mu} \Delta^{\nu} 
	\theta_{\mu \nu}^{GV} \, , \\ \nonumber
	& = & - \frac{1}{\alpha} \Delta \cdot A^a 
	\Delta \cdot \partial ( \partial \cdot A^a ) - ( \Delta \cdot 
	\partial \xi^a ) \Delta \cdot D^{ab} \omega^b \,.
\end{eqnarray}
The Feynman rules for $O_{em}^{GI}$ and $O_{em}^{GV}$ which we require 
are given in Appendix A.
One can show that $O_{em}^{GI}$ is invariant under the gauge transformation 
defined above and that $O_{em}^{GV}$ is BRST exact, meaning that 
it can be written 
as a BRST variation of another operator called its ancestor.  Explicitly,
\begin{equation}
O_{em}^{GV} = \delta_{BRST} \Omega \, , 
\end{equation}
with
\begin{equation}
\Omega = ( \Delta \cdot \partial \xi^a ) \Delta \cdot A^a \, ,
\end{equation}
as can be checked using (2.5) and (2.6).

In physical applications of perturbative QCD one is interested in the 
anomalous dimension of the twist two gauge invariant gluon operator \cite{gw}.
At zero momentum, contracted with $m$ powers of an 
arbitrary light like vector, $\Delta^{\mu}$, it is
\begin{equation}
O_{g}^{GI} (0) = \frac{1}{2} F_{\alpha}^{a_1} D^{a_1 a_2} \cdots 
D^{a_{m-2} a_{m-1}} F^{a_{m-1} \alpha}\,,
\end{equation}
where
\begin{eqnarray*}
F_{\alpha}^{a} & \equiv & \Delta^{\mu} F_{\alpha \mu}^a \, , \\
D^{a b} & \equiv & i \Delta^{\mu} D^{a b}_{\mu} \, ,
\end{eqnarray*}
and $m=2,4,6,\ldots$ \, .
One can show that $O_{g}^{GI}$ is invariant under the infinitesimal 
gauge transformation defined in (2.4).
We write $O_{g}^{GI}$ schematically as $F (D^{m-2} F)$.  As we intend 
to insert $O_{g}^{GI}$ at non-zero momentum the operator to be 
considered is
\begin{equation}
O_{g}^{GI} (Q) = F (D^{m-2} F) + (DF) (D^{m-3} F) + \cdots 
+ (D^{m-3} F) (DF) + (D^{m-2} F) F \, .
\end{equation}
One can show by explicit construction that it is possible to find 
a $\Delta^{\mu}$ 
such that $\Delta \cdot Q = 0$ but $Q=p_1-p_2 \neq 0$.  So that 
\begin{equation}
S_{O^{GI}_{g}} = \int d^4x e^{i x \cdot Q}  O_{g}^{GI} (Q) \,,
\end{equation}
can be integrated by parts and put in the form of the zero momentum case 
above, i.e. $F (D^{m-2} F)$.  For practical calculations this involves nothing 
more that using the relation $\Delta \cdot p_1 = \Delta \cdot p_2$.
For convenience, the normalizations of $O_{em}^{GI}$  and  $O_{g}^{GI}$ 
are chosen so that 
at $m=2$ they coincide.  The Feynman rules for $O_{g}^{GI}$ which we require 
are also given in Appendix A.

\section{General Results}
A large body of literature exists on the renormalization of composite 
operators in gauge theories.  Various gauges have been explored including
the axial, covariant and background field gauges.  In this paper we 
will only be concerned with the covariant gauge results.

The covariant gauge investigations began with the work of Gross and Wilczek 
\cite{gw} and culminated in a series of papers by Joglekar and Lee, and 
Joglekar \cite{jog}.  Most 
intermediate references are cited in \cite{jog} with the exception of 
\cite{clee} where the first order UV pole pieces of the energy-momentum 
tensor were calculated in the general covariant gauge and shown to add to 
zero after wave function renormalization.  This calculation was done for 
non-zero momentum insertion and showed that the 
energy-momentum tensor for pure unbroken Yang-Mills is finite to first order.  
However, no effort was made to retain the finite pieces (to see if they 
were gauge independent) nor were the physical matrix elements of 
the $O_{em}^{GV}$ operator calculated (to see if they vanish).  
Also of interest is the simplified proof of some parts 
of the Joglekar and Lee paper using cohomology techniques \cite{hen} and 
a recent detailed study of the energy-momentum tensor inserted at {\em zero} 
momentum \cite{randy} wherein Collins and Scalise  
isolate an IR divergence that spoils the proof of Theorem 2 below.

We begin with some terminology.
A non-zero momentum insertion means that there is a momentum flowing 
through the operator vertex.  For example, we take $p_1$ flowing out one 
leg, $p_2$ flowing in the other, and $Q=p_1-p_2$ flowing into the vertex 
as shown in Figure (1a).  
Gauge independent means that the result is independent of the gauge fixing 
parameter $\alpha_r$.  
Gauge invariant means that the operator is invariant 
under the gauge variation defined in Eq. (2.4).
A physical matrix element of an operator is an on-shell truncated proper 
(on-shell amputated one-particle-irreducible)
Green function with an insertion of an operator.
By on-shell we mean external legs are contracted with physical polarization 
vectors, put on-mass-shell, and properly renormalized.

We now quote without proof some of the theorems \cite{jog}, \cite{clee}, 
\cite{collins} on the renormalization 
of composite operators in non-Abelian gauge field theories.
As has been noted in \cite{jog}, \cite{clee}, and \cite{randy} these 
theorems only apply at non-exceptional momentum.  That is, the operators 
should be inserted at non-zero momentum.

Let $G_i$ be a set of gauge invariant operators which mix among 
themselves under renormalization, $A_i$ be the set of 
gauge variant operators which mix with $G_i$ under renormalization, and 
$E_i$ be the set of operators which vanish by use of the equations of motion 
which mix with $G_i$ under renormalization.

\newtheorem{theorem}{Theorem}

\begin{theorem}
$A_i$ can be chosen so that it is BRST exact up to equations of motion. i.e.
\begin{equation}
A_i \approx \delta_{BRST} B_i \, .
\end{equation}
\end{theorem}
\begin{theorem}
Physical matrix elements of BRST exact operators are zero.
\end{theorem}
\begin{theorem}
Physical matrix elements of $E_i$ are zero.
\end{theorem}
\begin{theorem}
Physical matrix elements of $G_i$ are independent of the gauge fixing 
parameter $\alpha_r$.
\end{theorem}
\begin{theorem}
The renormalization mixing matrix is triangular.  i.e.
\begin{equation}
\left( \begin{array}{c} R \left[ G \right] \\ R \left[ A \right] 
\\ R \left[ E \right] \end{array} \right) =
\left( \begin{array}{ccc} 
Z_{GG} & Z_{GA} & Z_{GE} \\
0      & Z_{AA} & Z_{AE} \\
0      & 0      & Z_{EE}
\end{array} \right)
\left( \begin{array}{c} G  \\ A \\ E \end{array} \right) .
\end{equation}
\end{theorem}
It is Theorems 2, 4 and 5, that, at first sight, appear to be violated 
by the work of Hamberg and van Neerven \cite{ham} who, following 
\cite{gw} and \cite{sac}, worked with zero momentum insertions.  
We demonstrate in the next section that these theorems 
hold to one loop when one considers non-zero momentum insertions and 
works in space-time dimension $n>4$.

\section{Explicit Calculations}
In this section we outline the calculation of the first order physical 
matrix element with two external gauge fields and an insertion of 
$O^{GI}_{em}$, $O^{GV}_{em}$, and $O_{g}^{GI}$ at non-zero momentum.
We do this in two ways.  First, we try a traditional LSZ prescription 
and a ``modified LSZ'' prescription \cite{randy} and see that these 
methods fail to respect Theorem 4 given above.  
Second, we work in $n>4$ space-time dimensions and consider truly 
massless external legs and find that the theorems given 
in the previous section are confirmed in our one loop example.
As mentioned above, the first order pole pieces of the 
matrix elements with an insertion of the energy-momentum 
tensor at non-zero momentum have been studied in 
detail in \cite{clee}.  In \cite{randy}, a detailed study of the 
energy-momentum tensor inserted at zero momentum was made.

We calculate matrix elements of operators inserted at
$Q=p_1-p_2 \neq 0$ thereby introducing another mass scale $Q^2$ into 
the problem.  The algebra is therefore more involved than 
in the case of $Q = 0$. The tensor integrals were reduced to scalar 
integrals as discussed in Appendix B.  For this process we performed 
all algebra on a computer using the program {\scriptsize FORM} \cite{jos}, 
except for the inversion of the large matrices involved, for which we used 
{\scriptsize MAPLE V} \cite{maple}. 
The Feynman rules for the vertices containing $O_{em}^{GI}$, $O_{em}^{GV}$, 
and $O_{g}^{GI}$ are given in Appendix A.  We used the gauge field propagator
\begin{equation}
D^{ab}_{\mu \nu}(k) = \frac{\delta^{ab}}{k^2} \left( g_{\mu \nu} - 
(1-\alpha_r) \frac{k_{\mu} k_{\nu}}{k^2} \right) \, ,
\end{equation}
in all the calculations which follow along 
with the other standard vertices from the effective Lagrangian.

\subsection{Calculating in $n=4$ dimensions}

In calculating physical matrix elements when truncating the 
external propagators one keeps a square root of the
residue of the pole of the full propagator for each leg 
as one goes on-shell.  The other
square root is discarded with the truncated propagator.  To find the
residue, note that the full propagator is \cite{muta}
\begin{equation}
\widetilde{D}^{ab}_{\mu \nu} (k^2) = \frac{\delta^{ab}}{k^2} \left[
\frac{g_{\mu \nu} - k_{\mu} k_{\nu} / k^2}{1+\Pi(k^2)} + \alpha_r
\frac{k_{\mu} k_{\nu}}{k^2} \right].
\end{equation}
The self energy of the YM field is well known \cite{self}.  It is 
derived by expanding terms $(-k^2)^{\epsilon/2}$ for small $\epsilon$, 
using $x^{\epsilon} = 1 + \epsilon \ln(x) + \cdots$, then 
after the contribution from the Lagrangian
counterterm is added the $1 /\epsilon$ terms cancel and 
$\epsilon$ is set to zero.  However, $k^2$ is not set to zero 
due to the presence of terms in $\ln(-k^2)$.  We write
\begin{equation}
\Pi_{\mu \nu}^{ab} (k) = \delta^{ab} (k_{\mu} k_{\nu} - k^2 
	g_{\mu \nu} ) \Pi(k^2) \, ,
\end{equation}
then
\begin{equation}
\Pi(k^2) = \frac{g_r^2}{16 \pi^2} C_A \left\{
        - \frac{31}{9} + \frac{5}{3} \ln (\frac{-k^2}{\mu^2})
        + (1-\alpha_r) \left[ 1 + \frac{1}{2} \ln (\frac{-k^2}{\mu^2}) \right]
        - \frac{1}{4} (1-\alpha_r)^2 \right\} \, ,
\end{equation}
where we have used the $\overline{MS}$ scheme.  The wave function 
renormalization constant in this scheme is
\begin{equation}
Z_3 = 1 - \frac{g_r^2}{16 \pi^2} C_A  \frac{1}{\bar{\epsilon}}
\left[ \frac{10}{3} + (1-\alpha_r) \right] .
\end{equation}
In the above equations 
${\bar{\epsilon}}^{-1} = \epsilon^{-1} + \frac{1}{2} \gamma
- \frac{1}{2} \ln 4 \pi$, $\epsilon = n-4$, 
$f^{acd} f^{bcd} = \delta^{ab} C_A$, $\gamma$
is the Euler constant, and we have extracted the dimension of the coupling 
by introducing the mass scale $\mu$.  
We are now in the position to extract the 
coefficient of the pole as $k^2 \rightarrow 0$.  It is normally chosen as 
\begin{eqnarray}
R & = & \left[ 1 + \Pi(k^2) |_{k^2 \rightarrow 0} \right]^{-1}
	\, , \\ \nonumber
  & = & 1 - \frac{g_r^2}{16 \pi^2} C_A \left\{
        - \frac{31}{9} + \frac{5}{3} \ln (\frac{-k^2}{\mu^2})
        + (1-\alpha_r) \left[ 1 + \frac{1}{2} \ln (\frac{-k^2}{\mu^2}) \right]
        - \frac{1}{4} (1-\alpha_r)^2 \right\} .
\end{eqnarray}
We will see below that if we use
\begin{eqnarray}
\widetilde{R} & = & \left\{ 1 + \frac{\partial [ k^2 \Pi(k^2)]}{\partial k^2} 
	|_{k^2 \rightarrow 0} \right\}^{-1} \, , \\ \nonumber 
	& = & 1 - \frac{g_r^2}{16 \pi^2} C_A \left\{
        - \frac{31}{9} + \frac{5}{3} (1+\ln (\frac{-k^2}{\mu^2}))
        + (1-\alpha_r) \left[ \frac{3}{2} + \frac{1}{2} \ln
        (\frac{-k^2}{\mu^2}) \right]
        - \frac{1}{4} (1-\alpha_r)^2 \right\} \, ,
\end{eqnarray}
as the residue, as suggested by Collins and Scalise \cite{randy}, 
then at $Q=0$ the physical matrix element of $O_{em}=O_{em}^{GI}+O_{em}^{GV}$ 
remains unaffected by higher order corrections and is independent of the 
gauge fixing parameter $\alpha_r$.  This absorbs infrared singularities 
into the asymptotic in and out states much like what one does in proofs of 
factorization in perturbative QCD.

\subsubsection{Energy-Momentum Tensor}

\paragraph{Gauge invariant operator, $O_{em}^{GI}$.}

\begin{figure}
\centerline{\hbox{
\psfig{figure=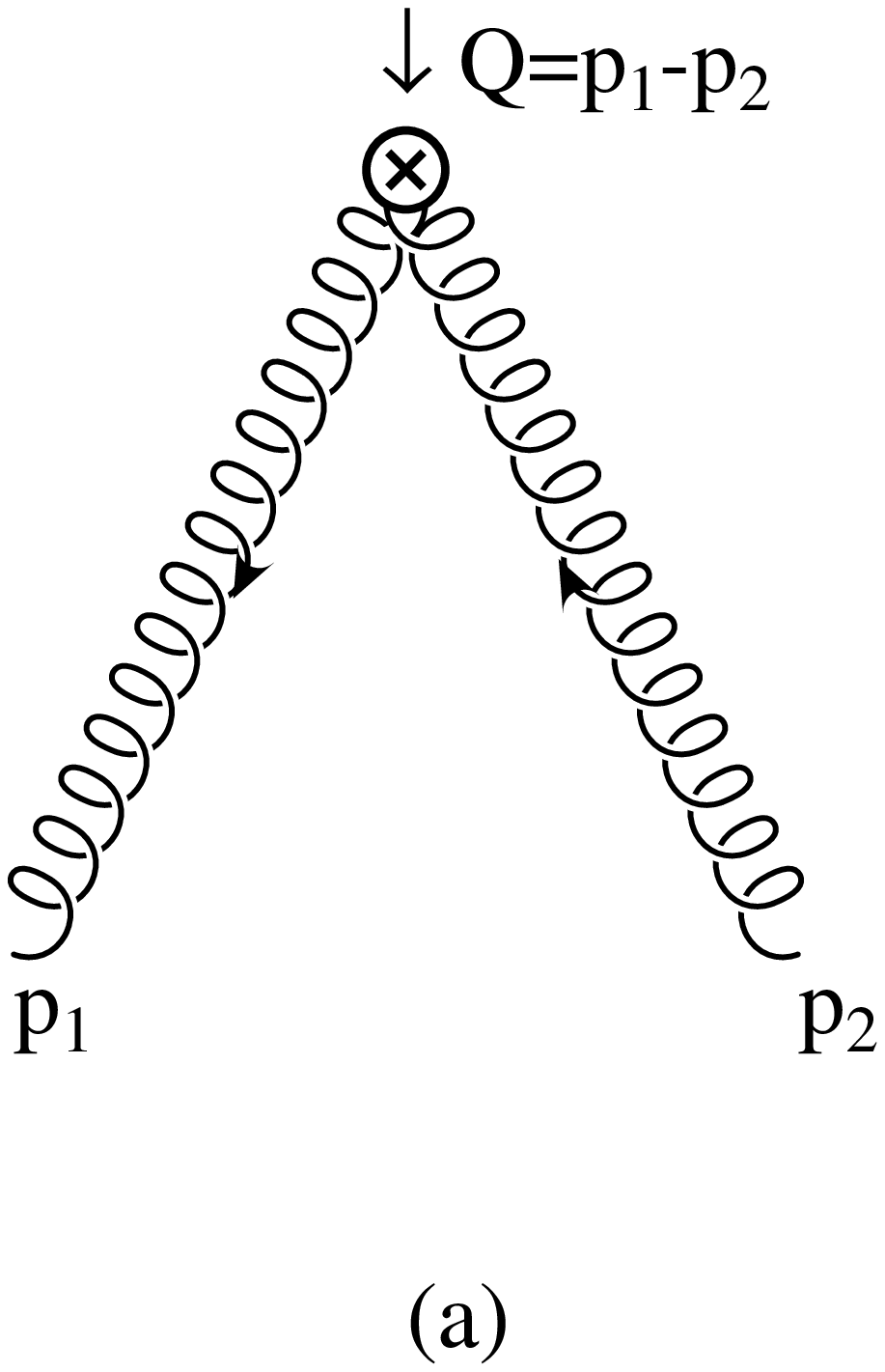,width=1in,height=1.5in}
\psfig{figure=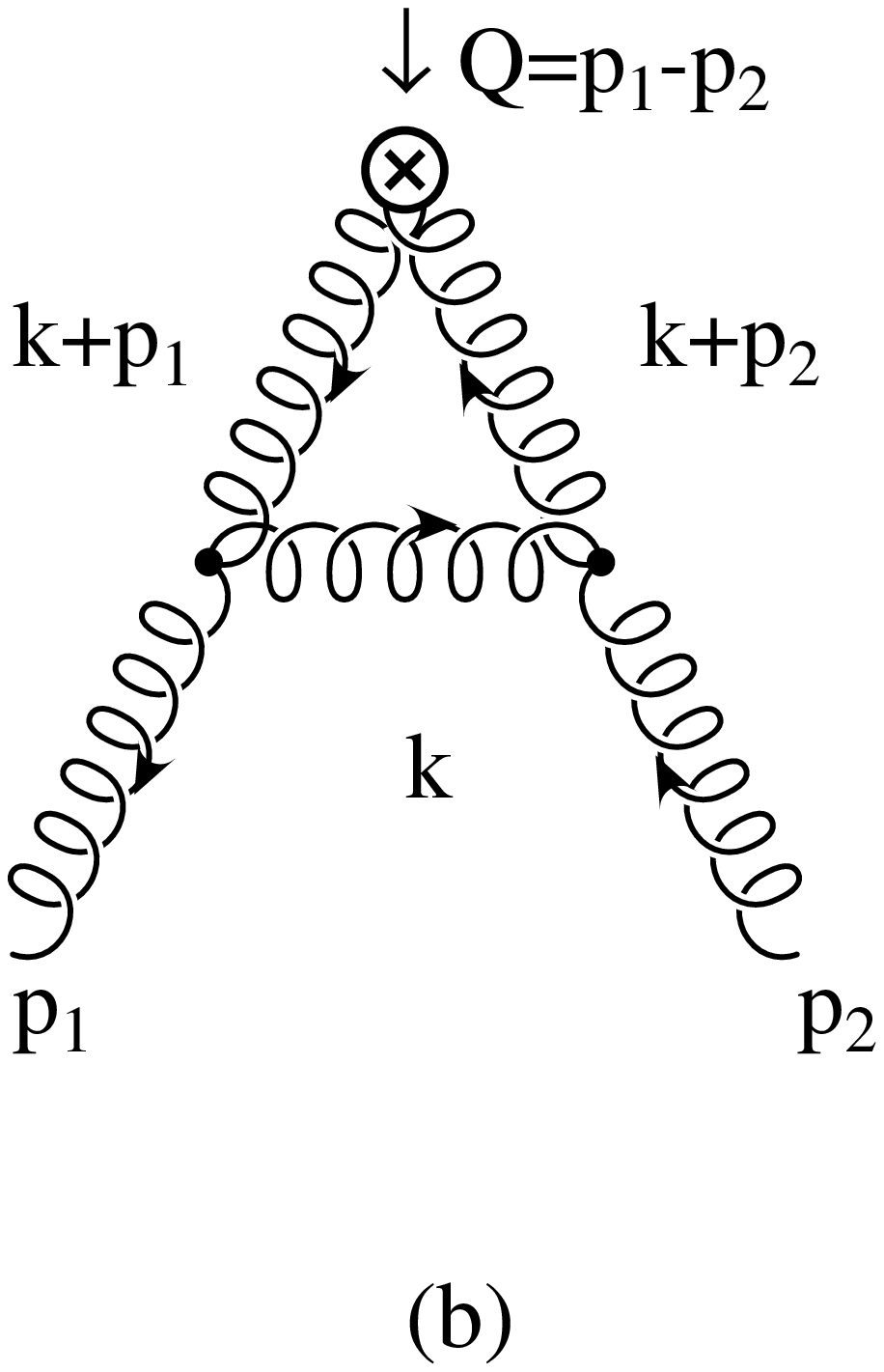,width=1in,height=1.5in}
\psfig{figure=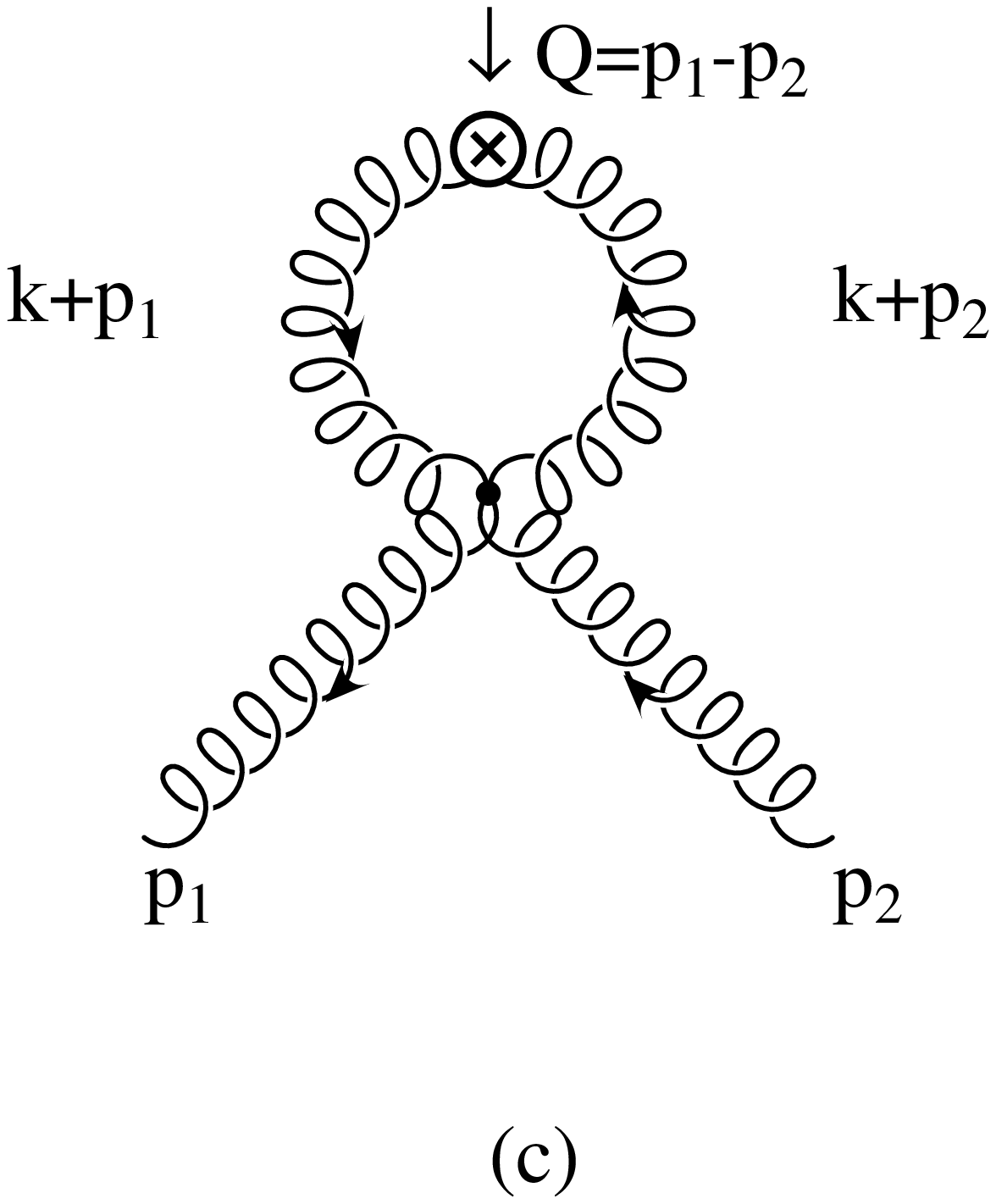,width=1in,height=1.5in}
\psfig{figure=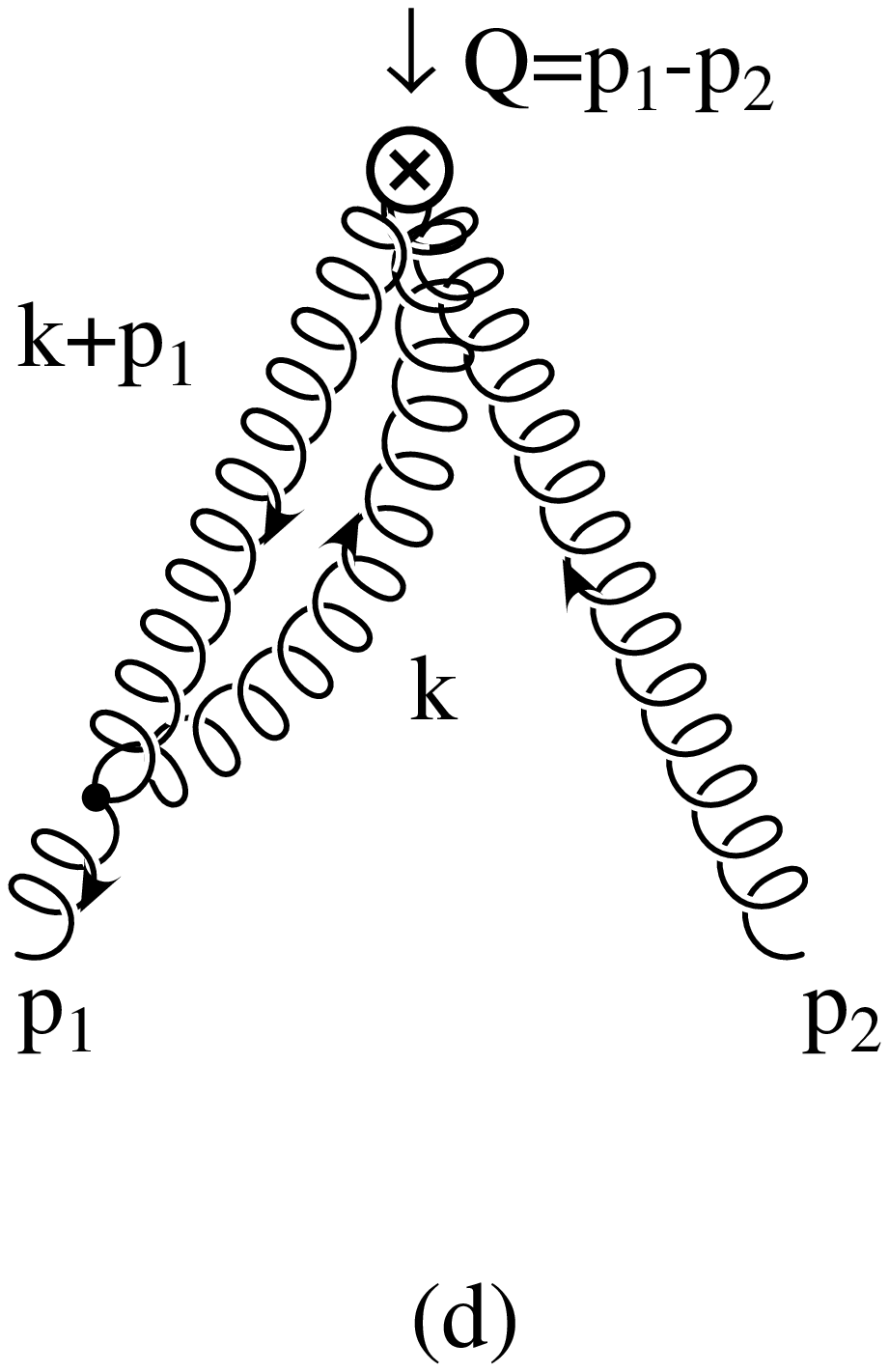,width=1in,height=1.5in}
\psfig{figure=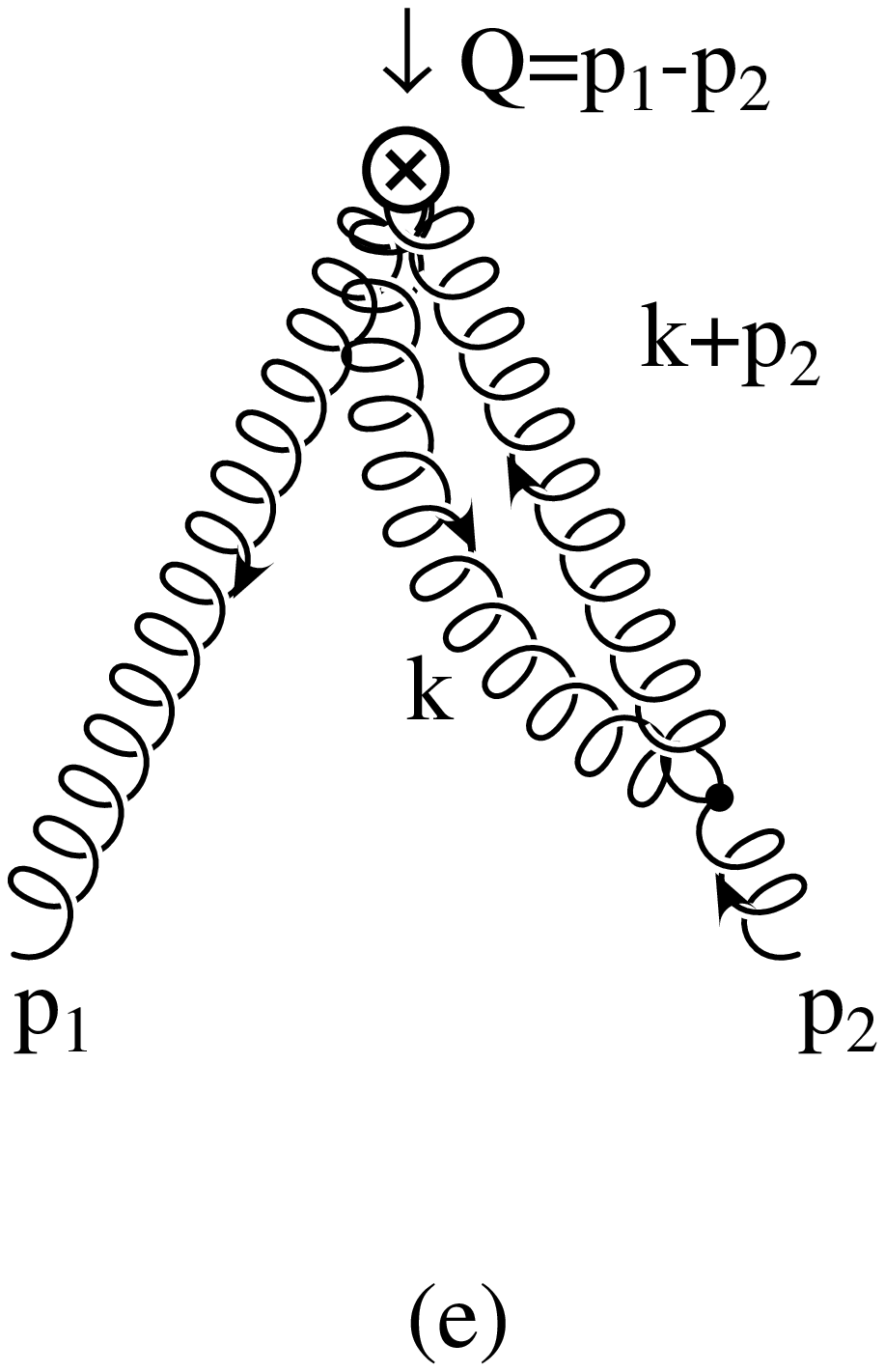,width=1in,height=1.5in}
\psfig{figure=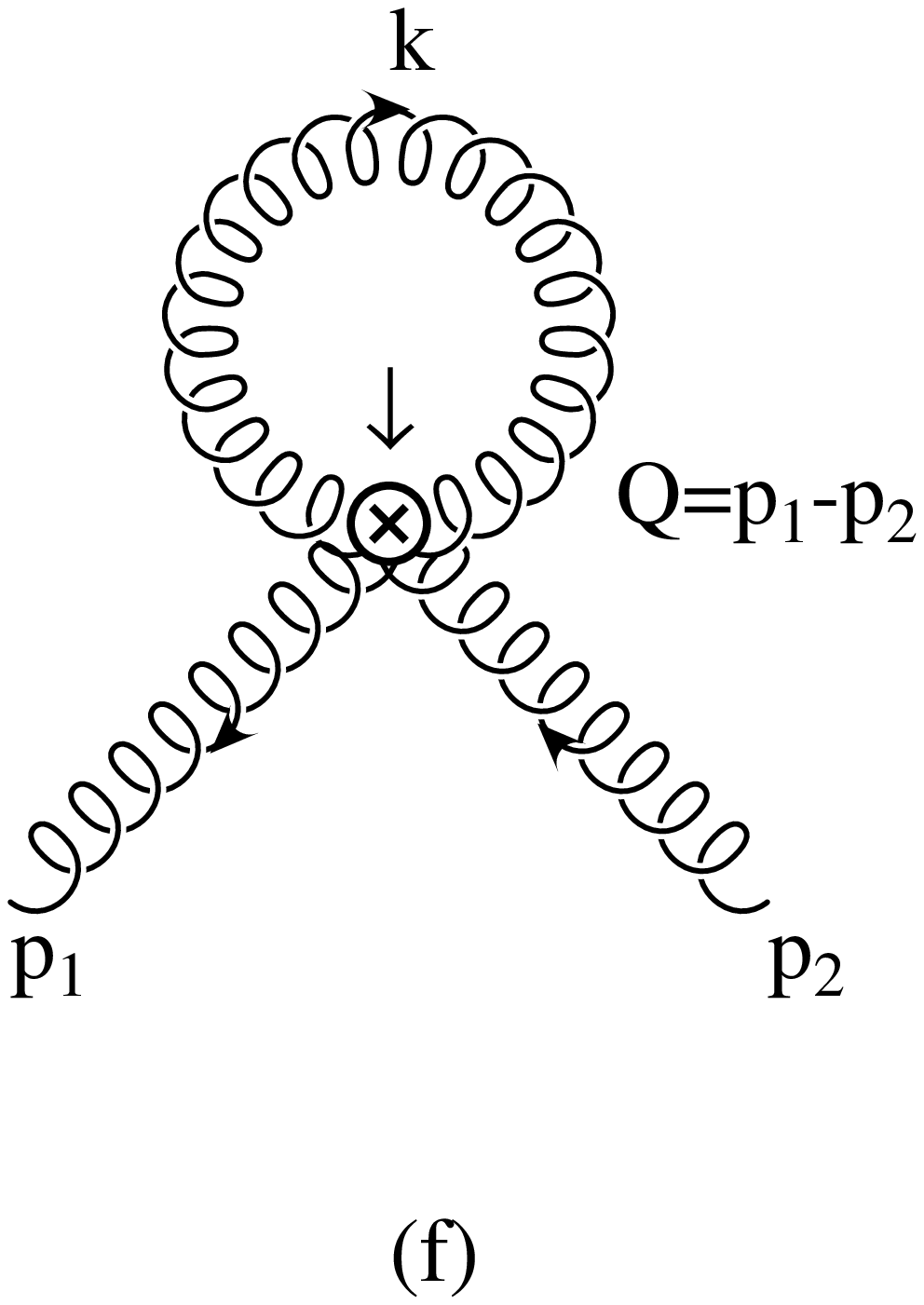,width=1in,height=1.5in}
}}
\caption{Diagrams contributing to the physical matrix element of 
$O_{em}^{GI}$ and $O_{g}^{GI}$ to one loop.  A circle with cross 
denotes an operator insertion.}
\end{figure}

We begin by considering the gauge invariant piece of the 
energy-momentum tensor.
The diagrams relevant for the calculation of the two leg 
physical matrix element at one loop are shown in Figure 1.  
The one loop amplitude $A_{\mu \nu}^{ab}(-p_1,p_2,2)$ is given by 
diagrams (b) - (e).  Diagram (f), which is constructed using the four 
leg $O_{em}^{GI}$ vertex 
is zero in dimensional regularization so it is not considered.
We were careful to include the symmetry factor $1/2$ for diagrams (c), 
(d), and (e) and then partial fraction the resulting expression to remove all terms 
of the form $k \cdot p_1$ and $k \cdot p_2$.  Next we contracted  with 
polarization vectors and used the on-shell conditions 
$\epsilon (p_1) \cdot p_1 = 0 = \epsilon (p_2) \cdot p_2$ and 
$p_1^2=p_2^2=m^2$ but refrained from setting $m^2=0$.

As a check, we set $Q=0$, performed 
a reduction to scalar integrals, and expanded the resulting expression 
in powers of $\epsilon = n-4$. The physical matrix element is,
\begin{eqnarray}
\label{GIatzero}
<O_{em}^{GI}(0)> & = &  \widetilde{R} Z_3 \left[ T^{ab}_{\mu \nu}(-p,p,2) 
	+ A^{ab}_{\mu \nu}(-p,p,2) \right] \epsilon^{\mu}(p) 
	\epsilon^{\nu}(p) \, , \\ \nonumber
	    & = & - \delta^{ab} (\Delta \cdot p)^2 + \delta^{ab} 
\frac{g_r^2}{16 \pi^2} C_A  (\Delta \cdot p)^2 \left[ 3 - 
\frac{3}{2} (1-\alpha_r) + \frac{1}{4} (1-\alpha_r)^2 \right] \, ,
\end{eqnarray}
where we have used $p_1=p_2 \equiv p$, $\epsilon(p) \cdot p = 0$, 
$p^2 =0$, and $\epsilon(p) \cdot \epsilon(p) = -1$.
If we would have used $R$ instead of $\widetilde{R}$ the coefficients of 
$ \{ 1,(1-\alpha_r),(1-\alpha_r)^2 \} $ would have been $ \{ 4/3,-2,1/4 \}$.
This agrees with \cite{ham} and \cite{randy} and provides a check 
on our calculation to this point.

Having made this check we now proceed with the $Q \neq 0$ calculation.  
The only difference is that the algebra and integrals become more involved.
A brief discussion of the methods is found in Appendix B.
We quote our findings for the physical matrix element of $O_{em}^{GI}$ as follows:
\begin{eqnarray}
\label{GIatnonzero1}
<O_{em}^{GI}(Q)> & = & \widetilde{R} Z_3 \left[ 
	T^{ab}_{\mu_1 \mu_2}(-p_1,p_2,2) 
	+ A^{ab}_{\mu_1 \mu_2}(-p_1,p_2,2) \right] \epsilon^{\mu_1}(p_1) 
	\epsilon^{\mu_2}(p_2) \, , \\ \nonumber
        & = & \epsilon^{\mu_1}(p_1) \epsilon^{\mu_2}(p_2) 
T_{\mu_1 \mu_2}^{ab} (-p_1,p_2,2) \\ \nonumber
	&  \times & \left( 1 - \frac{g_r^2}{16 \pi^2} C_A \left\{ 
	\frac{11}{2} - \frac{11}{3} \ln \left( \frac{Q^2}{m^2} \right)  
	+ 2 \ln^2 \left( \frac{Q^2}{m^2} \right) \right. \right. 
\\ \nonumber 
	& + & \left. \left. (1-\alpha_r) \left[ \frac{1}{2} - \ln
\left( \frac{Q^2}{m^2} \right) \right] + \frac{1}{4} (1-\alpha_r)^2 \right\} 
\right) \\ \nonumber	
	& + & \frac{g_r^2}{16 \pi^2} C_A \left[ \epsilon (p_1) \cdot 
\epsilon (p_2) - \frac{p_1 \cdot \epsilon (p_2) p_2 \cdot \epsilon (p_1)}
{p_1 \cdot p_2} \right] \\ \nonumber 
	& & \times \left[ \frac{2}{3} ( \Delta \cdot p_1 )^2 
- ( \Delta \cdot p_1 ) ( \Delta \cdot p_2 )  + \frac{2}{3} 
( \Delta \cdot p_2 )^2 \right] .
\end{eqnarray}
It is clear that this expression is ill defined at $m^2=0$ (so there is 
a problem with infrared divergences) and is dependent on the gauge fixing 
parameter $\alpha_r$.

\paragraph{Gauge variant operator, $O_{em}^{GV}$.}

\begin{figure}
\centerline{\hbox{
\psfig{figure=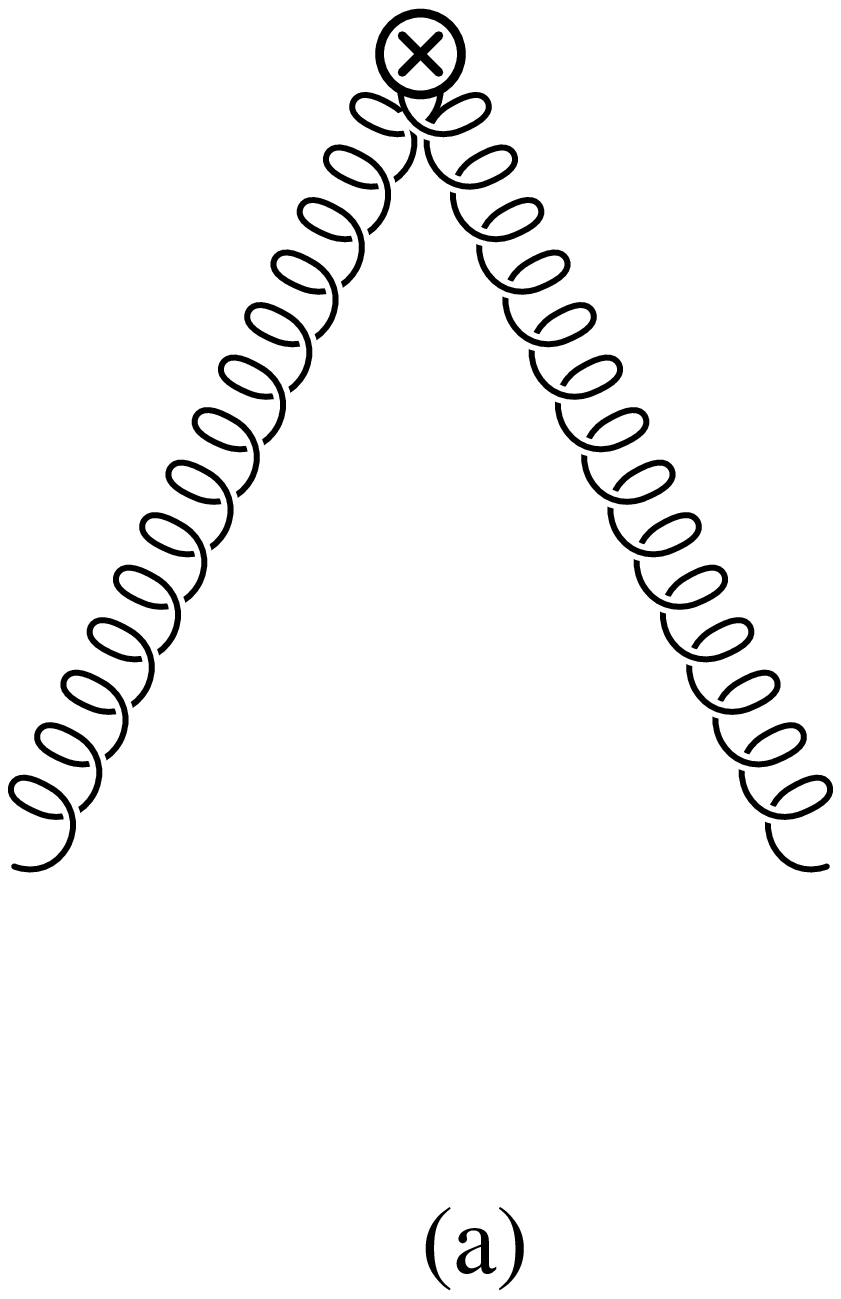,width=1in,height=1.5in}
\psfig{figure=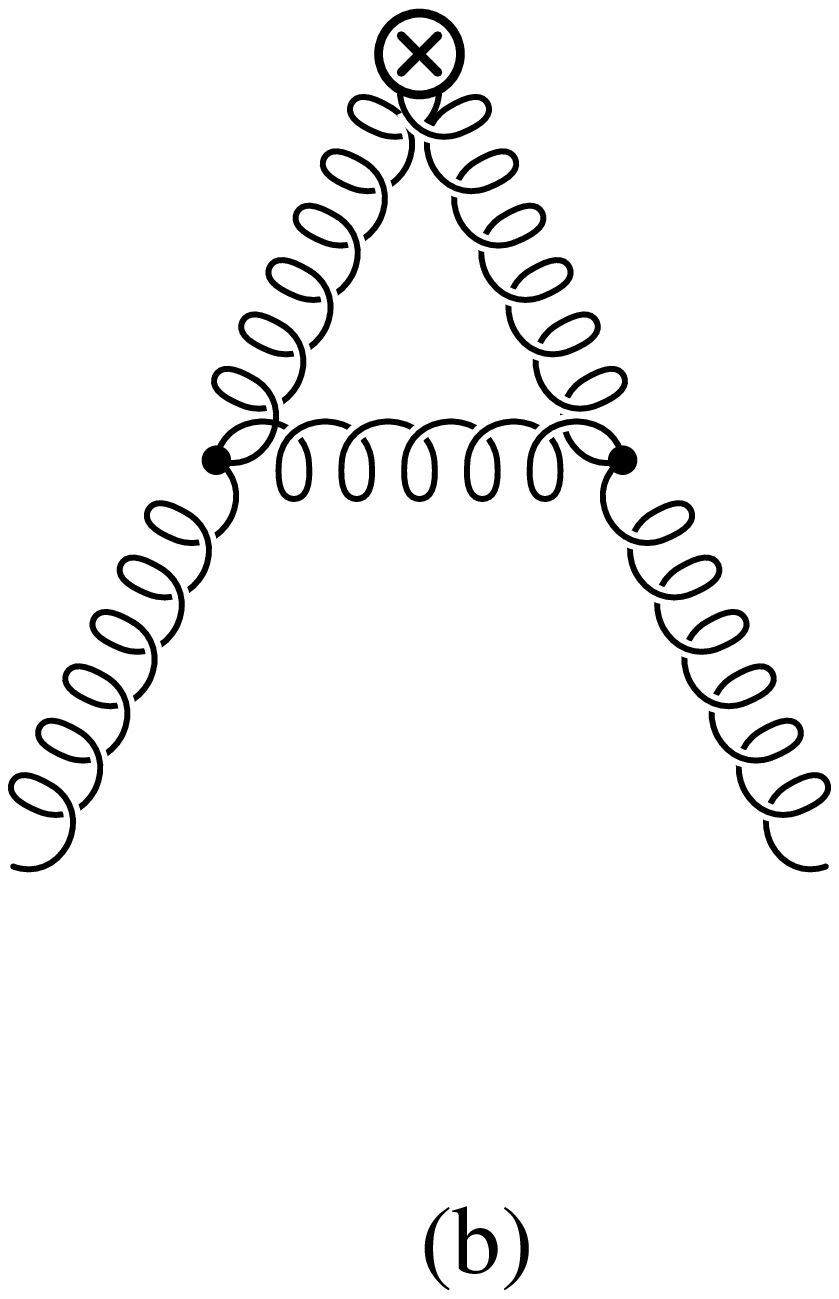,width=1in,height=1.5in}
\psfig{figure=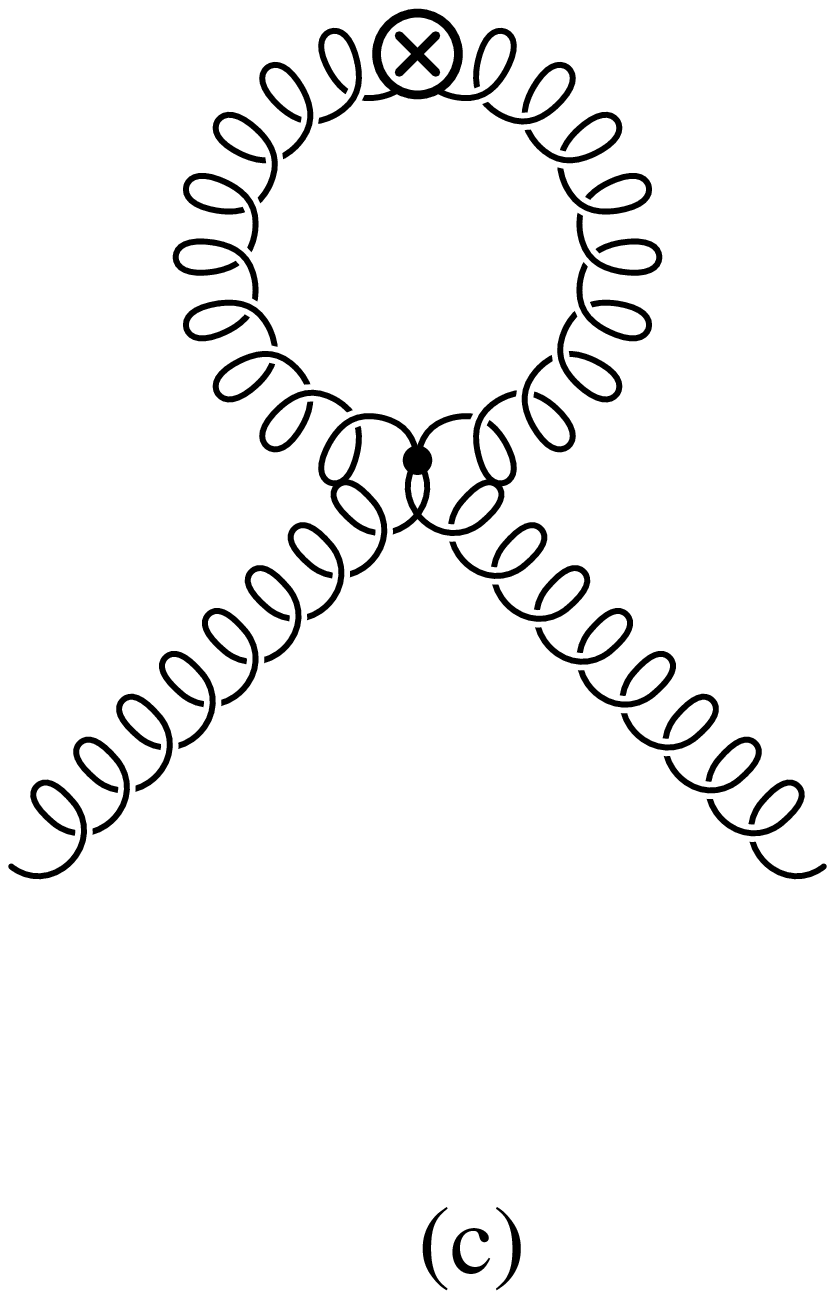,width=1in,height=1.5in}
\psfig{figure=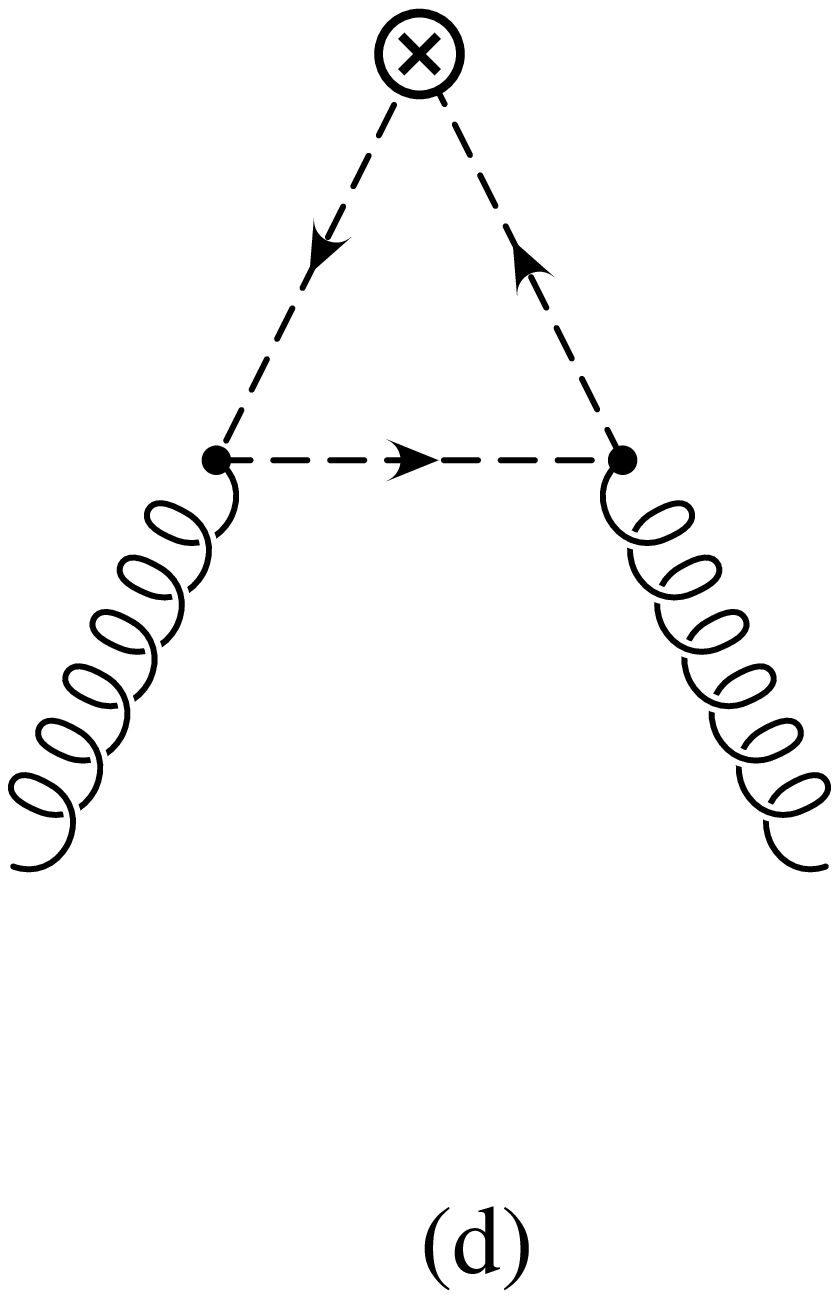,width=1in,height=1.5in}}}
\vskip 0.5cm
\centerline{\hbox{
\psfig{figure=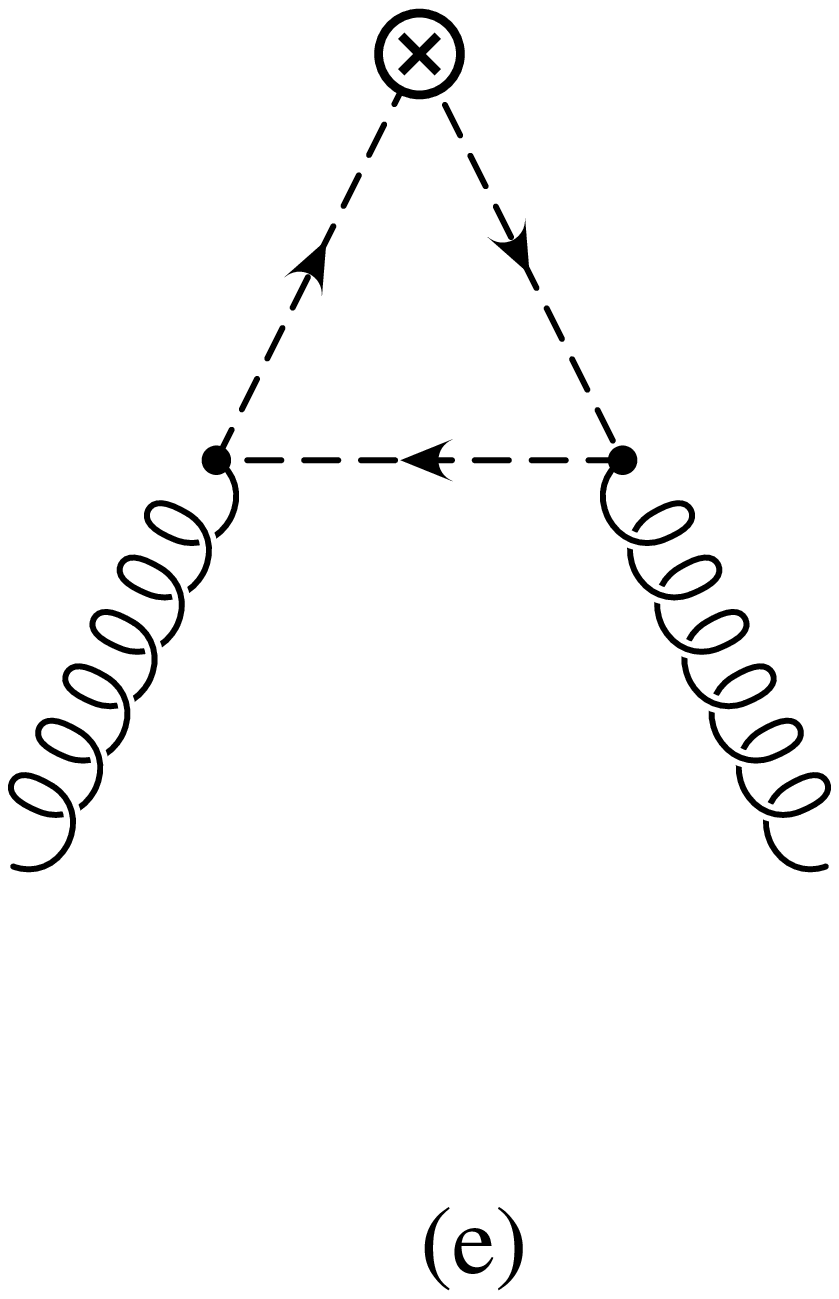,width=1in,height=1.5in}
\psfig{figure=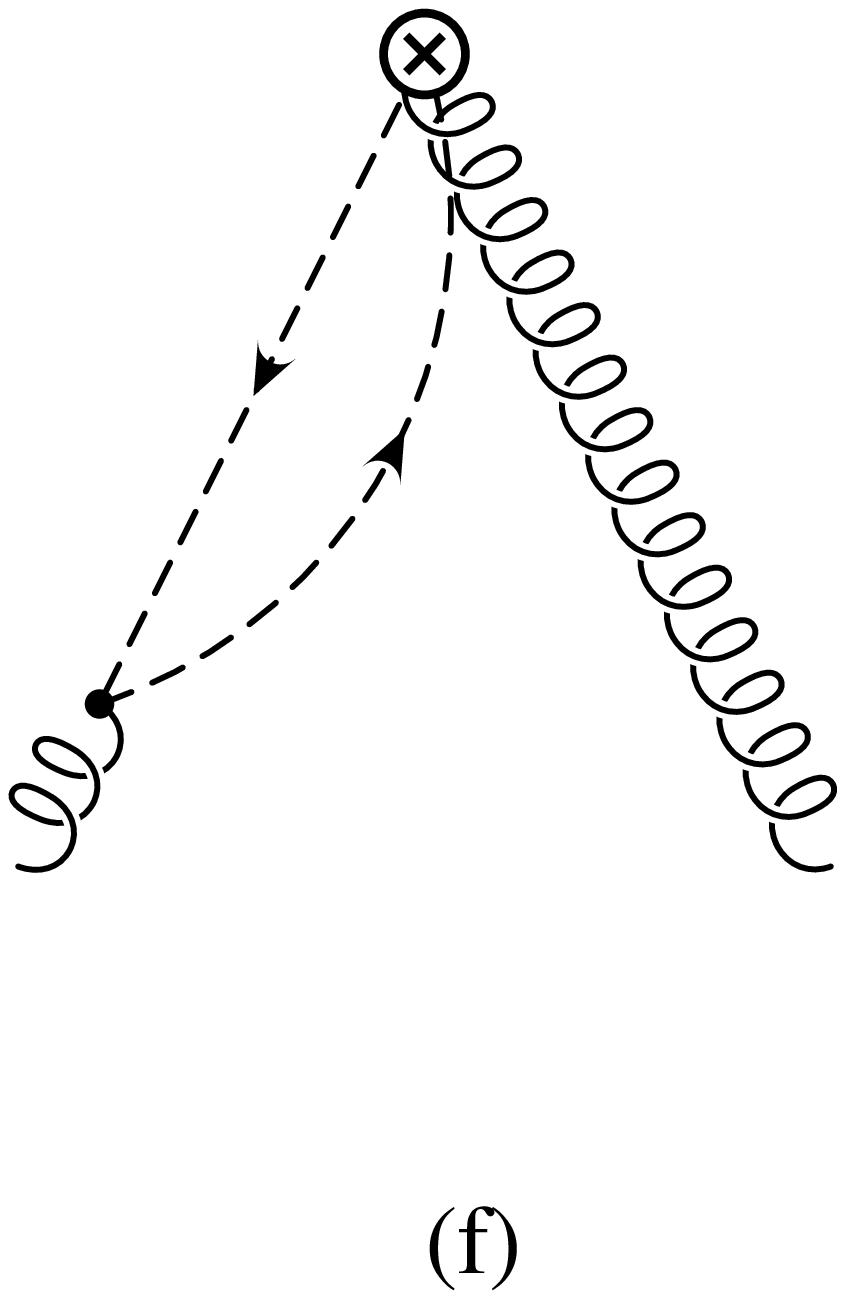,width=1in,height=1.5in}
\psfig{figure=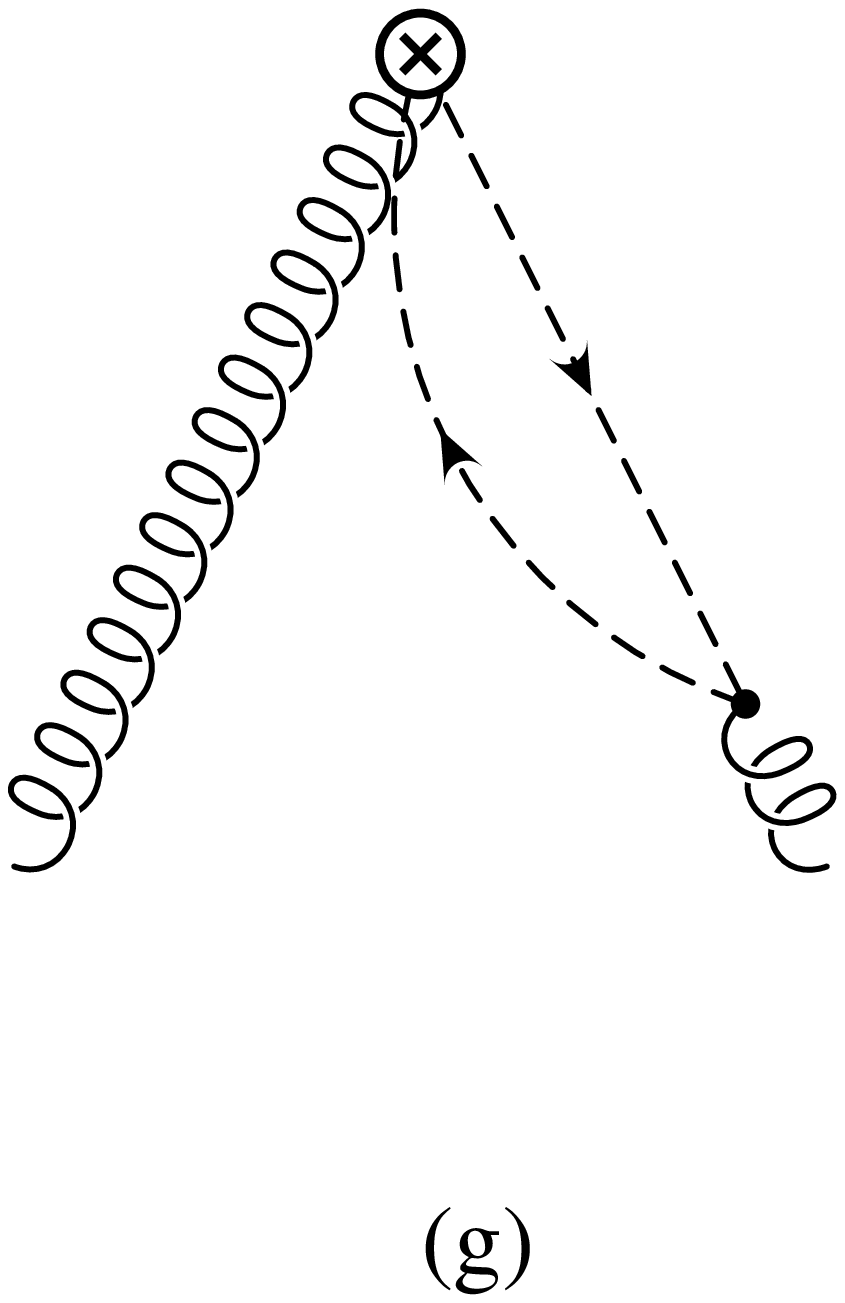,width=1in,height=1.5in}
}}
\caption{Diagrams contributing to the physical matrix element of 
$O_{em}^{GV}$ to one loop.  Momenta routings are the same as Figure 1.}
\end{figure}

We now consider the gauge variant piece of the energy-momentum tensor.
The diagrams relevant for the calculation of the two leg
physical matrix element at one loop are shown in Figure 2.  
The momenta are routed similar to those in Figure 1.
The tree level $O_{em}^{GV}$ diagram (a), does not contribute as it vanishes 
when contracted with polarization vectors.  
The one loop amplitude $B_{\mu \nu}^{ab}(-p_1,p_2)$ is the sum of 
diagrams (b)-(g) including the symmetry factor $1/2$ 
for diagram (c).  We partial fractioned the resulting expression 
and used the on-shell conditions as was done for the GI case.

At this point, as a check, we set $Q=0$, performed
a reduction to scalar integrals, and expanded the resulting expression
in powers of $\epsilon = n-4$. The physical matrix element is,
\begin{eqnarray}
\label{GVatzero}
<O_{em}^{GV}(0)> & = & \widetilde{R} Z_3 \left[ V_{\mu \nu}^{ab}(-p,p)
	+ B_{\mu \nu}^{ab}(-p,p) \right] \epsilon^{\mu}(p) 
	\epsilon^{\nu}(p) \, , \\ \nonumber 
	& = &  B_{\mu \nu}^{ab}(-p,p) \epsilon^{\mu}(p)  \epsilon^{\nu}(p) 
	\, , \\ \nonumber 
	& = & - \delta^{ab} \frac{g_r^2}{16 \pi^2} C_A (\Delta \cdot p)^2 
	\left[ 3 - \frac{3}{2} (1-\alpha_r) + \frac{1}{4} (1-\alpha_r)^2 
	\right] \, ,
\end{eqnarray}
where we have used $p_1=p_2 \equiv p$, $\epsilon(p) \cdot p = 0$, $p^2=0$ 
and $\epsilon(p) \cdot \epsilon(p) = -1$.  This again agrees with \cite{randy} 
and gives
\begin{equation}
<O_{em}(0)>  = <O_{em}^{GI}(0)> + <O_{em}^{GV}(0)> = - \delta^{ab} ( \Delta \cdot p )^2 .
\end{equation}
The form of the physical matrix element of $O_{em}$ is unaffected by 
higher order corrections and independent of the gauge fixing parameter 
with the use of $\widetilde{R}$ as the residue instead of $R$. 

We now proceed with the $Q$ non-zero result.  The only difference is that 
the algebra and integrals become more involved.  For the physical matrix 
element with $O_{em}^{GV}$ inserted and $Q$ non-zero we find zero.  Explicitly,
\begin{eqnarray}
\label{GVatnonzero1}
<O_{em}^{GV}(Q)> & = & \widetilde{R} Z_3 \left[ V_{\mu_1 \mu_2}^{ab}(-p_1,p_2)
	+ B_{\mu_1 \mu_2}^{ab}(-p_1,p_2) \right] \epsilon^{\mu_1}(p_1) 
	\epsilon^{\mu_2}(p_2) \, , \\ \nonumber
	& = &  B_{\mu_1 \mu_2}^{ab}(-p_1,p_2) \epsilon^{\mu_1}(p_1)  
	\epsilon^{\mu_2}(p_2) \, , \\ \nonumber
	& = & 0.
\end{eqnarray}
Which is in agreement with the Theorem 2 and is not affected by 
what we choose for $R$ because the tree level diagram 
$V_{\mu_1 \mu_2}^{ab}(-p_1,p_2)$  vanishes when contracted with 
polarization vectors satisfying $\epsilon(p) \cdot p = 0$.

By looking at Eq. (\ref{GIatzero}) we see that Theorem 4 is violated so 
at this point we already notice a problem.
By looking at Eq. (\ref{GVatzero}) we see that Theorem 2 is also violated.  
However, Eq. (\ref{GVatnonzero1}) confirms Theorem 2 but Eq. 
(\ref{GIatnonzero1}) has IR divergences in the form $\ln^i (Q^2/m^2), i=1,2$ 
that must be removed.  They are due to the order of our limits, 
namely, we took the limit $\epsilon \rightarrow 0$ too soon.  

Since there is an IR divergence we can extract it by reworking the 
calculation and taking the limits $p_1^2 \rightarrow 0$ and 
$p_2^2 \rightarrow 0$ before $\epsilon \rightarrow 0$.
In dimensional regularization, with space-time dimension $n=4+\epsilon$, 
both the UV and IR divergences are regulated. 
After UV renormalization, which is performed
at $\epsilon < 0$  we can analytically continue to $\epsilon > 0$
which regulates the IR singularities.  This allows us to 
set the square of the momenta of the external gluon lines 
to zero because $(p_1^2)^{\epsilon/2} = 0 = (p_2^2)^{\epsilon/2}$ so 
long as $\epsilon > 0$.  This interchange of the order of the limits 
is the crux of our new calculation which we present next.

\subsection{Calculating in $n > 4$ dimensions}

In contrast to the previous section, here we reverse the order of limits and 
find that Theorem 4 is now satisfied.  That is, we consider 
$\epsilon > 0$ so that $(-k^2)^{\epsilon/2} = 0$ in the limit $k^2=0$.  
This implies that the residue factors $R Z_3$ of the previous section are 
to be set to unity.  We will see that doing this leads to a universal double 
and single pole structure that allows use to determine the anomalous 
dimension of the gluon operator.

\subsubsection{Energy-momentum tensor revisited}

\paragraph{Gauge invariant operator, $O_{em}^{GI}$.}

We now return to the gauge invariant piece of the energy-momentum tensor.
The diagrams relevant for the calculation of the two particle 
physical matrix element at one loop are shown in Figure 1.  
Diagram (f), which is constructed using the four leg $O_{em}^{GI}$ 
vertex, is zero in dimensional regularization.  Also, because 
we set $(p_1^2)^{\epsilon/2} = 0 = (p_2^2)^{\epsilon/2}$ 
diagrams (d) and (e) vanish.  We partial fractioned the expression 
resulting from diagrams (b) and (c) to remove all terms 
of the form $k \cdot p_1$ and $k \cdot p_2$.  Next we contracted  with 
polarization vectors $\epsilon^{\mu_1}(p_1)$ and  $\epsilon^{\mu_2}(p_2)$ 
and used the on-shell conditions 
$\epsilon (p_1) \cdot p_1 = 0 = \epsilon (p_2) \cdot p_2$ and 
$p_1^2=0=p_2^2$ remembering that
$(p_1^2)^{\epsilon/2} = 0 = (p_2^2)^{\epsilon/2}$ as $\epsilon >0$.
A discussion of the methods for evaluating the resulting integrals 
is found in Appendix B.
Adding diagram (a) we quote our findings for the sum of all
the diagrams in Figure 1 with $O_{em}^{GI}$ inserted at $Q$ and 
external legs on-shell as defined above as follows:
\begin{eqnarray}
\label{GIatnonzero2}
<O_{em}^{GI}(Q)> & = & 1 \left[ T^{ab}_{\mu_1 \mu_2}(-p_1,p_2,2) 
	+ A^{ab}_{\mu_1 \mu_2}(-p_1,p_2,2) \right] \epsilon^{\mu_1}(p_1) 
	\epsilon^{\mu_2}(p_2) \, , \\ \nonumber
	 & = & \epsilon^{\mu_1}(p_1) \epsilon^{\mu_2}(p_2) 
	T_{\mu_1 \mu_2}^{ab} (-p_1,p_2,2) \\ \nonumber
	&  \times & \left\{ 1 + g_r^2 C_A S_n (-\frac{Q^2}{\mu^2})
	^{\epsilon}\left[ 
	-\frac{8}{\epsilon^2} + \frac{-4 \gamma_E + 22/3}{\epsilon}
	+ h(2) \right] \right\} \\ \nonumber	
	& + & g_r^2 C_A S_n (-\frac{Q^2}{\mu^2})
	^{\epsilon} \left[ \epsilon (p_1) \cdot 
	\epsilon (p_2) - \frac{p_1 \cdot \epsilon (p_2) p_2 
	\cdot \epsilon (p_1)} {p_1 \cdot p_2} \right] \\ \nonumber 
	& & \times \left[ \frac{2}{3} ( \Delta \cdot p_1 )^2 
	- ( \Delta \cdot p_1 ) ( \Delta \cdot p_2 )  + \frac{2}{3} 
	( \Delta \cdot p_2 )^2 \right] \, ,
\end{eqnarray}
where $C_A = N_c = 3$, $S_n = \pi^{n/2} / (2 \pi)^n$, and 
$T^{ab}_{\mu_1 \mu_2}(-p_1,p_2,2)$ is the Born level vertex for the 
classical energy-momentum tensor given explicitly 
in Appendix A.  The factor of 
$\mu^{-2 \epsilon}$ is the dimension of the coupling $g_r$.
The finite term $h(2)$ is given later in Eq. (\ref{h}) for completeness.

At this point we pause to note several important things.  One, this result 
is independent of the gauge fixing parameter in agreement with Theorem 4.  
Two, because we are considering on-shell legs we have not multiplied by 
an LSZ type residue when truncating the legs.  Three,  there are both double 
and single Sudakov IR poles.  Four, there is a new finite form factor that was 
not present at the Born level.  Five, if we take the limit $Q^2 \rightarrow 0$ 
for $\epsilon > 0$ the only piece that survives is the Born level vertex.
We will return to these points in the discussion below.

\paragraph{Gauge variant operator, $O_{em}^{GV}$.}

We now return to the gauge variant piece of the energy-momentum tensor.
The diagrams relevant for the calculation of the corresponding
physical matrix element with an insertion of $O_{em}^{GV}$ 
at one loop order are shown in Figure 2.
The momenta are routed similar to those in Figure 1.
Again, (f) and (g) vanish.
The tree level $O_{em}^{GV}$ diagram (a), does not contribute as it vanishes 
on-shell.  After performing the integrations as above the result vanishes,
which is in agreement with the Theorem 2.  Explicitly,
\begin{eqnarray}
\label{GVatnonzero2}
< O_{em}^{GV} (Q) >  & = & 1 \left[ V_{\mu_1 \mu_2}^{ab}(-p_1,p_2)
	+ B_{\mu_1 \mu_1}^{ab}(-p_1,p_2) \right] \epsilon^{\mu_1}(p_1) 
	\epsilon^{\mu_2}(p_2) \, , \\ \nonumber 
	& = & B_{\mu_1 \mu_2}^{ab}(-p_1,p_2) \epsilon^{\mu_1}(p_1) 
	\epsilon^{\mu_2}(p_2) \, , \\ \nonumber
	& = & 0.
\end{eqnarray}

\subsubsection{Gluon operator}

Now consider the twist two gauge invariant gluon operator.
The one loop diagrams are shown in Figure 1.  
We followed the same procedure as outlined above for the revisited 
energy-momentum tensor but this time we took the liberty of setting 
$\Delta \cdot p_1 = \Delta \cdot p_2 \equiv \Delta \cdot p$ which allows 
us to do the integration by parts described at the end of Section II.  The 
necessary integrals are discussed in Appendix B.
We quote the result for the sum of all diagrams on-shell as defined above:
\begin{eqnarray}
\label{Ogatnonzero}
<O_{g}^{GI}(Q)>  & = & 1 \left[ T^{ab}_{\mu_1 \mu_2}(-p_1,p_2,m) 
	+ A^{ab}_{\mu_1 \mu_2}(-p_1,p_2,m) \right] \epsilon^{\mu_1}(p_1) 
	\epsilon^{\mu_2}(p_2) \, , \\ \nonumber
	& = & \epsilon^{\mu_1}(p_1) \epsilon^{\mu_2}(p_2) 
	T_{\mu_1 \mu_2}^{ab} (-p_1,p_2,m) \\ \nonumber
	& \times & \left\{ 1
	+ g_r^2 C_A S_n (-\frac{Q^2}{\mu^2})^{\epsilon}\left[ 
	-\frac{8}{\epsilon^2} + \frac{-4 \gamma_E + 22/3 
	+ \gamma^{(m)}_{gg}/C_A}{\epsilon} + h(m)  \right] \right\} 
\\ \nonumber	
	& + & g_r^2 C_A S_n (-\frac{Q^2}{\mu^2})^{\epsilon} (\Delta \cdot p)^m 
	f(m) \left[ \epsilon (p_1) \cdot \epsilon (p_2) 
	- \frac{p_1 \cdot \epsilon (p_2) p_2 \cdot \epsilon (p_1)}
	{p_1 \cdot p_2} \right] \,,
\end{eqnarray}
where $T_{\mu_1 \mu_2}^{ab} (-p_1,p_2,m)$ is the Born level vertex for 
the gauge invariant gluon operator given in Appendix A.  All other symbols are 
as defined below Eq. (\ref{GIatnonzero2}) and $\gamma^{(m)}_{gg}$ 
is the one loop anomalous dimension of the gluon operator \cite{gw}
\begin{equation}
\gamma^{(m)}_{gg} = C_A \left( \frac{8}{m+2} - \frac{8}{m+1} + \frac{16}{m} 
	- \frac{8}{m-1} + 8 \sum_{i=1}^{m-1} \frac{1}{i} - \frac{22}{3}
	\right).
\end{equation}
The finite functions $h(m)$ and $f(m)$ are given by
\begin{eqnarray}
\label{h}
h(m) & = & -\gamma_E^2+\frac{\gamma_E}{2} \left( 
	\frac{\gamma_{gg}^{(m)}}{C_A} + \frac{22}{3} \right) 
	+ \frac{\pi^2}{6} \\ \nonumber
     & - & 8 \widetilde{S}(m-2) - 8 S_1(m-1) 
	\left( \frac{1}{m+2} - \frac{1}{m+1}  
	+ \frac{2}{m} \right)  \\ \nonumber
     & - & \frac{8}{(m+2)^2} + \frac{12}{m+2} + \frac{8}{(m+1)^2} 
	- \frac{16}{m+1} - \frac{16}{m^2} + \frac{4}{m} \, ,
\end{eqnarray}
and
\begin{equation}
f(m) = 4 \left( \frac{-1}{m+2} + \frac{1}{m+1} \right) \, ,
\end{equation}
where
\begin{equation}
S_i(n) = \sum_{l=1}^{n} \frac{1}{l^i} \, ,
\end{equation}
\begin{equation}
\widetilde{S}(n) = \sum_{l=1}^{n} \left( \begin{array}{c} n 
	\\ l \end{array} \right)\frac{(-1)^{l+1}}{l^2} 
	= \frac{1}{2} S_2(n-1) + \frac{1}{n} S_1(n) + \frac{1}{2} 
	S_1^2(n-1) \, .
\end{equation}

The same points we noted below Eq. (\ref{GIatnonzero2}) apply here 
and are discussed below.  In addition, we note that the coefficient 
of the double pole is the same as that in Eq. (\ref{GIatnonzero2}), i.e. it 
is universal, and that the coefficient of the single pole contains the 
anomalous dimension of the gluon operator in addition to the terms in 
Eq. (\ref{GIatnonzero2}).

\section{Discussion and Conclusions}
We now return to the three points of contention listed in the 
introduction.  There are two reasons why the matrix elements of the 
GV operators used by Hamberg and van Neerven \cite{ham} 
did not vanish.  One is that their operators are not BRST exact.  
Therefore,  Theorem 2 does not apply.  
One can see this by simply checking that at Born level their $O_a$ 
does not vanish on-shell when $Q \neq 0$.
Even if they had used an operator that was BRST exact 
they would have found it had a non-vanishing physical matrix element because 
Theorem 2 breaks down when the operator is inserted at zero momentum as 
was shown in detail by Collins and Scalise \cite{randy}.  
The fact that Theorem 2 breaks 
down at one loop for a zero momentum insertion means that at two loops 
one will see a non-triangular mixing matrix, hence the apparent demise of 
Theorem 5.  By inserting BRST exact operators at non-zero momentum we 
have shown Theorem 2 works at one loop and we therefore have 
no reason to doubt Theorem 5.

The gauge parameter dependence of the physical 
matrix element of the gauge invariant 
gluon operator was not investigated by the first group 
attempting a two loop calculation in the covariant gauge \cite{sac}.  
They worked in the Feynman gauge.  Hamberg and van Neerven noticed 
this gauge dependence as they calculated the $O(\alpha_s)$ terms in the 
general covariant gauge.  
They claimed that this implied a renormalization of the gauge parameter in 
the two loop calculation to keep the coefficients of the single pole 
terms gauge independent.
They then proceeded with the two loop calculation in the Feynman gauge.  
Hence, they can not make a statement on the gauge parameter dependence 
of the finite terms at two loops.  
A finite gauge parameter dependence at two loops can lead to 
$1 / \epsilon$ gauge parameter dependence at three loops.  This would 
imply a gauge dependent anomalous dimension.  
As we demonstrated, our result is 
independent of the gauge fixing parameter at one loop.  
It is now safe to proceed to two loops.

In our method we put the momentum squared of the external legs to zero at 
the very beginning.  This is possible because we work in $n > 4$ dimensions 
and only after extracting the IR safe anomalous dimension do we let 
$n \rightarrow 4$.  
Because we worked with physical legs the LSZ residue is unity.
We also notice a double and single Sudakov pole structure arising 
because we set the momentum squared of the external legs to zero in 
the very beginning.  We conjecture that these factors are universal to all 
three point functions involving two external gluon legs.  An all orders proof 
of the universality may be possible and a two loop calculation of 
the energy-momentum tensor is technically possible and would serve as a 
check on these ideas.  
A detailed study of the universality of double and single pole structure will 
also yield information on the finite form factor seen in 
Eq. (\ref{GIatnonzero2}) and Eq. (\ref{Ogatnonzero}).
As can be seen from Eq. (\ref{GIatnonzero2}) it is safe to let 
$Q^2 \rightarrow 0$ as $\epsilon > 0$ (only IR divergences remain) 
leaving only the Born level result.  The same is true for Eq. 
(\ref{Ogatnonzero}) after the UV counter term has been added.

The fact that it took so long to verify the general theorems was 
anticipated in \cite{clee} where, following the proofs of the general 
theorems, there are statements like
\begin{quote}
\ldots matrix elements of $\tilde{N}(x)$ vanish if the on-shell limit 
can be taken. \ldots but for pure Yang-Mills theory, the on-shell limit 
cannot be taken because of infrared divergences \ldots
\end{quote}
It is only by seeing that these infrared divergences 
are universal if regulated 
dimensionally that we are able to overcome these problems and verify the 
general theorems.  Our method of calculation has actually been applied to 
the two loop anomalous dimension of the nonsinglet quark operator \cite{mats} 
where the complications of GI and GV operator mixing do not enter and where 
the $\alpha_s C_F$ and $\alpha_s^2 n_f C_F$ terms were 
found to be in agreement with the Sudakov factorization theorems 
\cite{sudakov}.

To conclude, we have presented new one loop calculations that vindicate 
the general theorems of Joglekar and Lee 
which were recently called into question 
by the work of Hamberg and van Neerven.  In the process we have found a 
new method of calculating the anomalous dimension of the gluon operator.  
We did this by considering physical matrix elements with insertions of 
the appropriate operator at {\em non-zero} momentum with the resulting 
IR singularities being regulated dimensionally.  
We found that the physical matrix element of either the classical 
energy-momentum 
tensor or the gauge invariant gluon operator is independent of the 
gauge fixing parameter.  
We also found that a physical matrix element of the BRST exact gauge variant 
operator that enters the full energy-momentum tensor vanishes.
A universal Sudakov factor appears in both the physical matrix 
elements of the
energy-momentum tensor and of the gluon operator.  The universality of 
this factor and the UV finiteness of the energy-momentum tensor then 
provide another method for determining the gluon anomalous dimension which 
we conjecture applies to higher loops as well.

\vspace{0.75in}
\centerline{\bf{Acknowledgements}}
\vspace{0.5in}
The authors thank John Collins and Randy Scalise for many stimulating 
discussions.  We also acknowledge George Sterman for a helpful discussion and 
Willy van Neerven and Peter van Nieuwenhuizen for comments.
The work in this paper was supported in part 
by the contract NSF 9309888.

\newpage


\appendix
\section{Feynman Rules}

\begin{figure}
\centerline{\hbox{
\psfig{figure=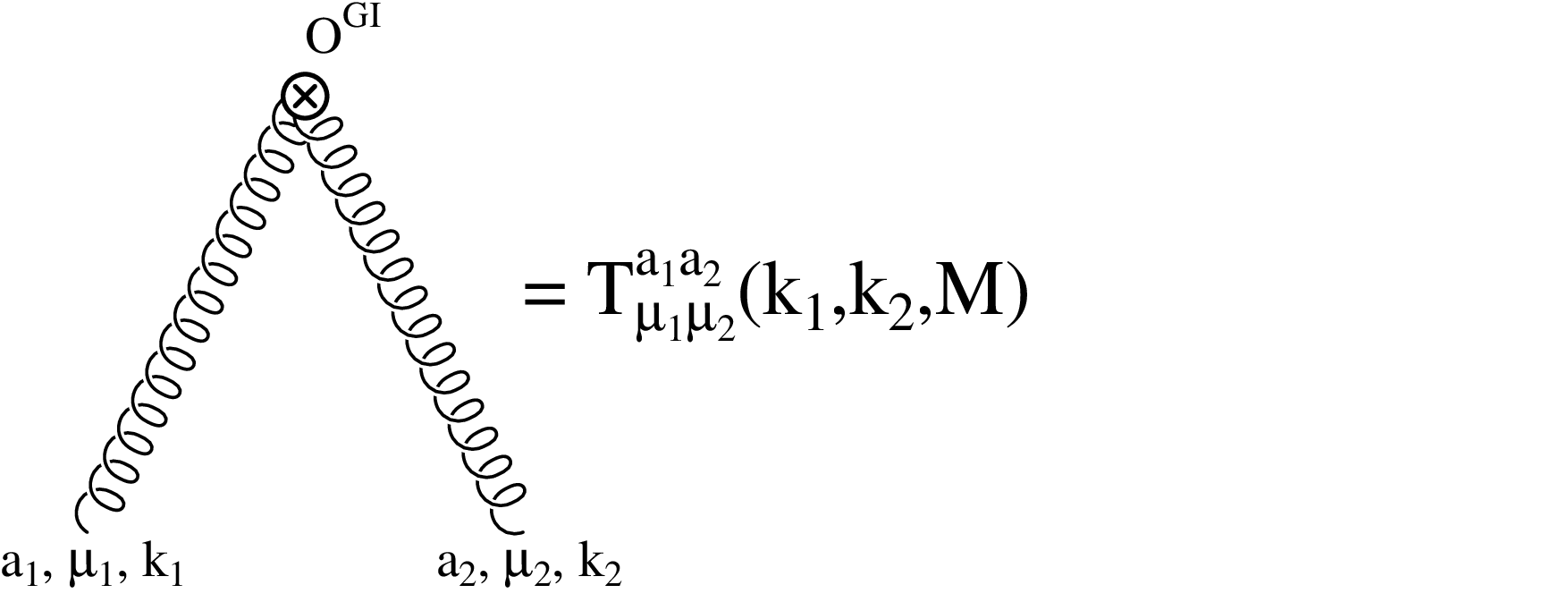,width=2.9in,height=1.2in}
\psfig{figure=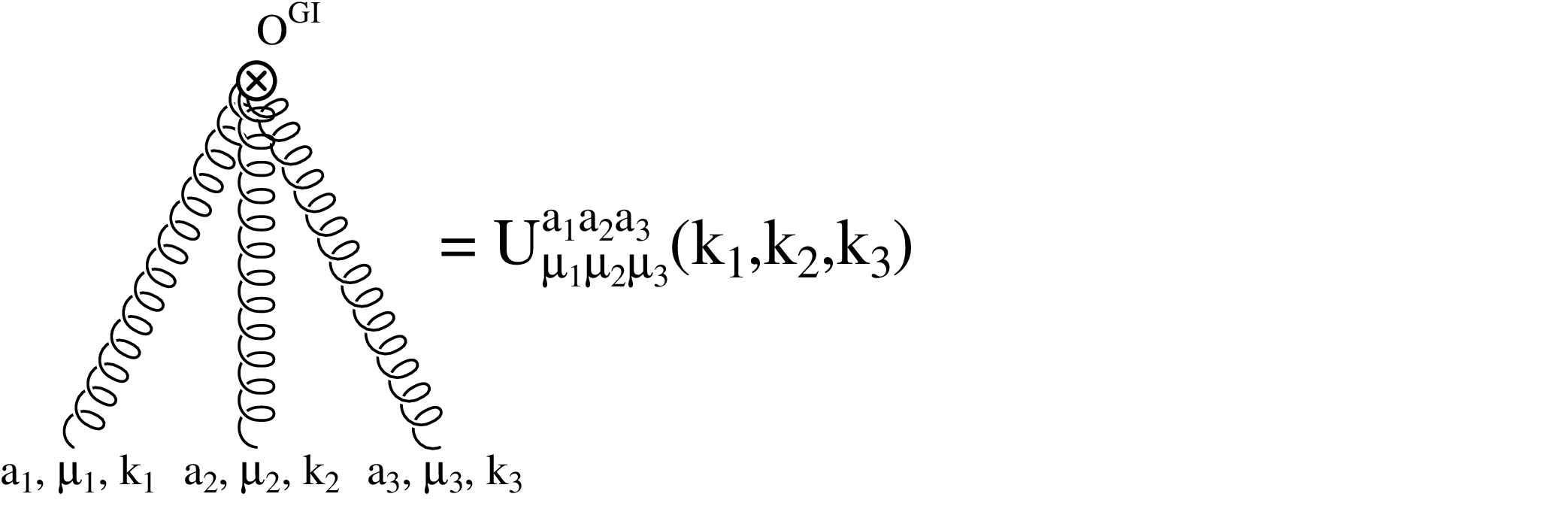,width=3.3in,height=1.2in}
}}
\caption{Feynman rules for $O_{em}^{GI}$ and $O_g^{GI}$}
\end{figure}

\begin{figure}
\centerline{\hbox{
\psfig{figure=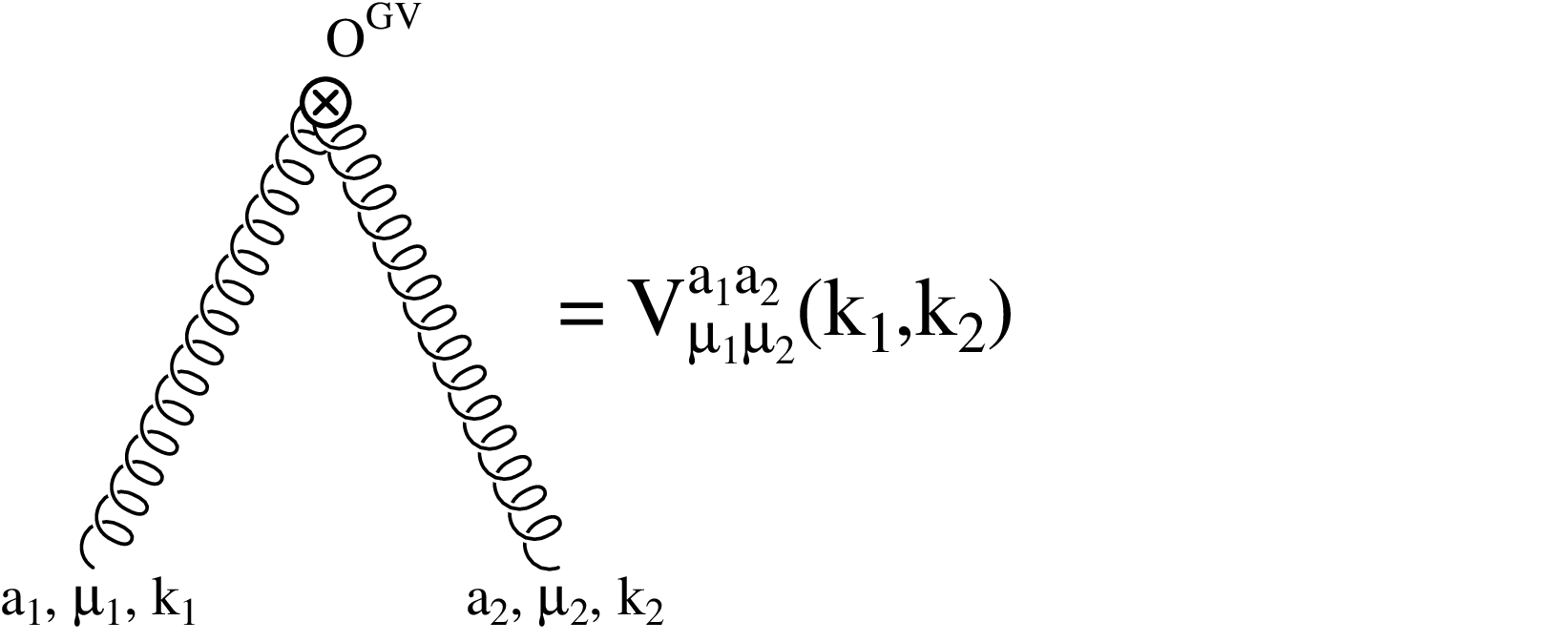,width=2.0in,height=1.2in}
\psfig{figure=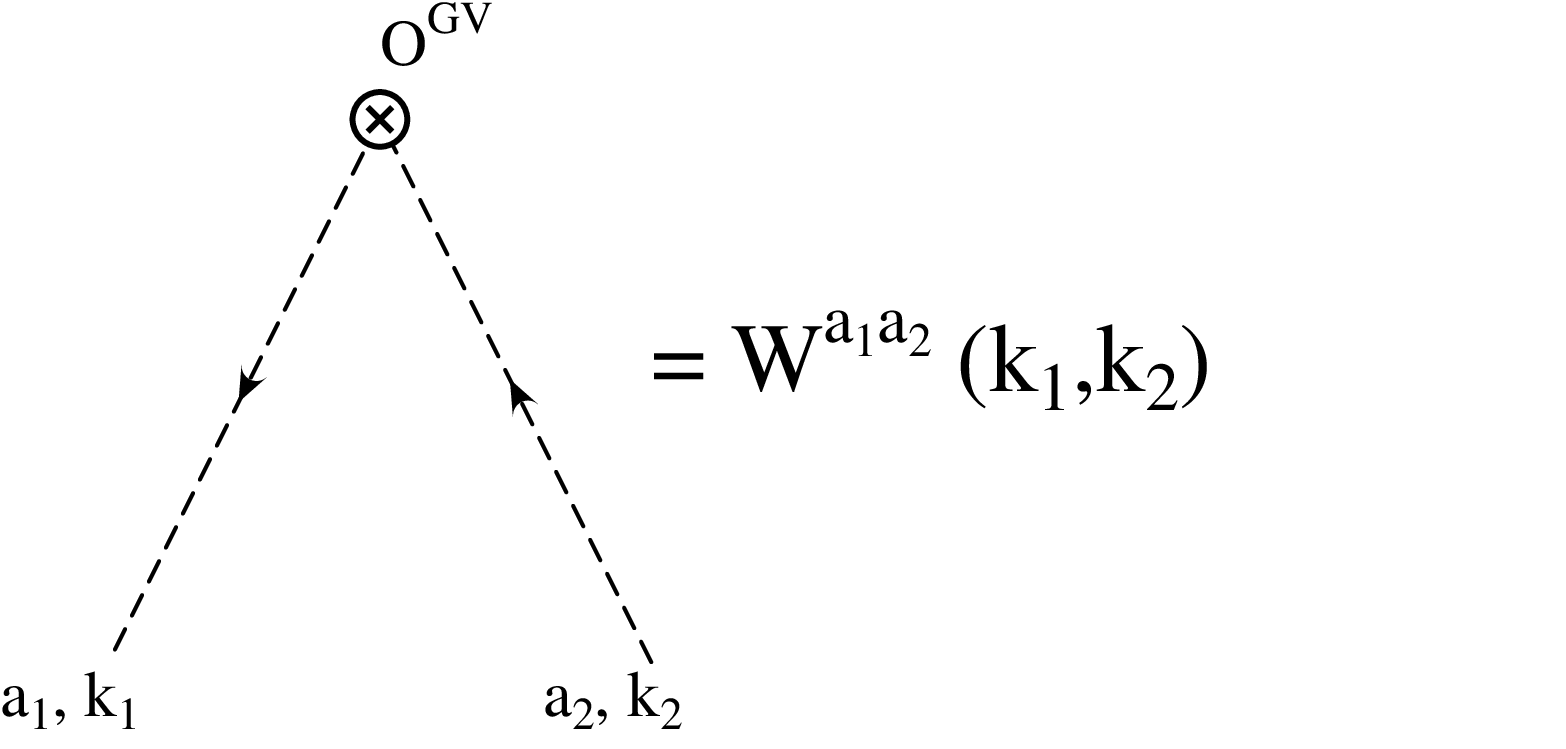,width=2.0in,height=1.2in}
\psfig{figure=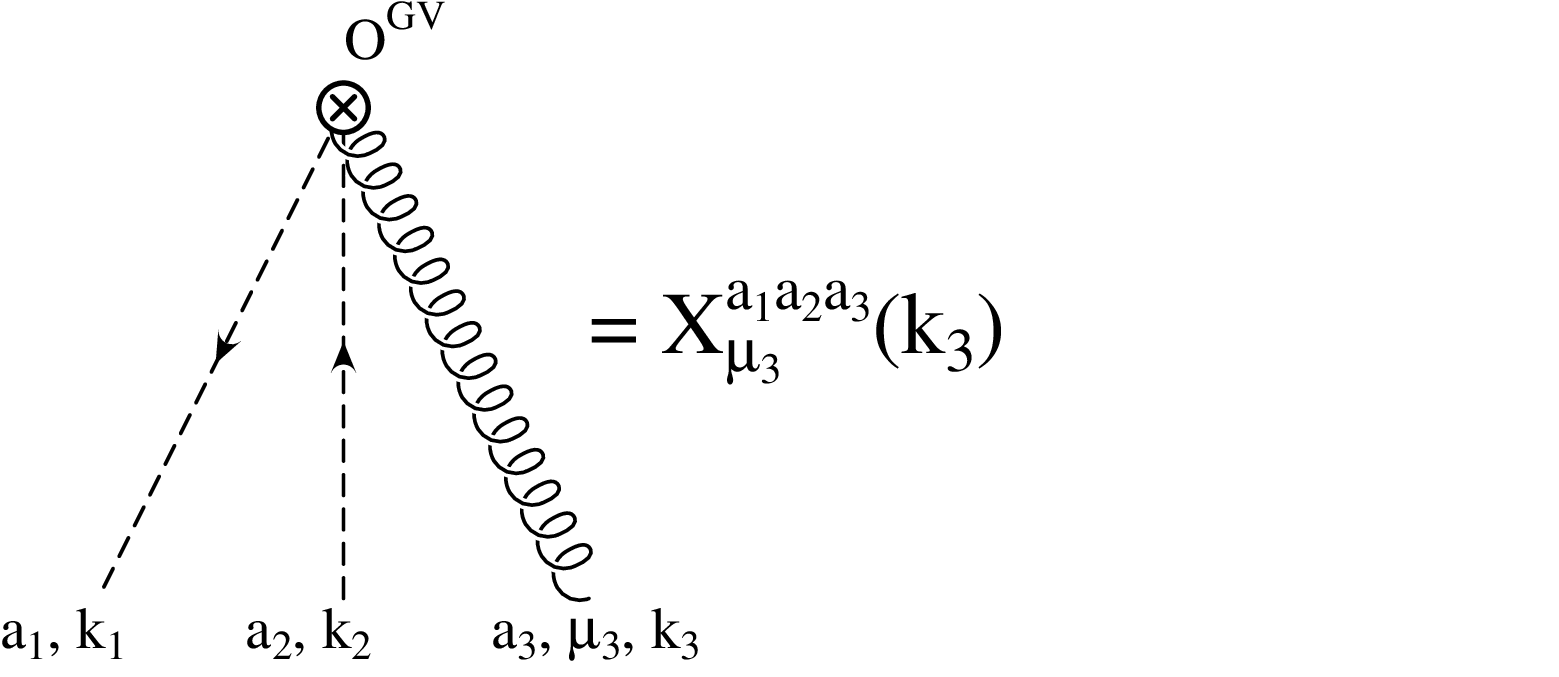,width=2.0in,height=1.2in}
}}
\caption{Feynman rules for $O_{em}^{GV}$}
\end{figure}

The operators $O_{em}^{GI}$, $O_g^{GI}$ and $O_{em}^{GV}$ have the 
following Feynman rules as shown in Figures 3 and 4.  All momenta 
flow into the vertices.
\begin{eqnarray}
T_{\mu_1 \mu_2}^{a_1 a_2} (k_1,k_2,M) & = & - \frac{1}{2} \delta^{a_1 a_2} 
	\left[ ( \Delta \cdot k_1 )^{M-2} + ( \Delta \cdot k_2 )^{M-2} 
	\right] \\ \nonumber 
	& \times & \left[ \Delta_{\mu_1} \Delta_{\mu_2} \, k_1 \! 
	\cdot \! k_2 - \Delta_{\mu_1} k_{1\mu_2} \, \Delta \! 
	\cdot \! k_2 \right. \\ \nonumber
        & - & \left. \Delta_{\mu_2} k_{2\mu_1} \, \Delta \! \cdot \! k_1
        + \Delta \! \cdot \! k_1  \, \Delta \! \cdot \! k_2 \,
        g_{\mu_1 \mu_2} \right] \, , \\
U_{\mu_1 \mu_2 \mu_3}^{a_1 a_2 a_3} (k_1,k_2,k_3)
        & = & - i g f^{a_1 a_2 a_3} \\ \nonumber 
        &  \times &        \left[ \Delta_{\mu_1}
                ( k_{1\mu_2} \Delta_{\mu_3} - \Delta_{\mu_2} k_{1\mu_3} )
                + \Delta \! \cdot \! k_1 ( g_{\mu_1 \mu_3} \Delta_{\mu_2}
                - g_{\mu_1 \mu_2} \Delta_{\mu_3} )  \right. \\ \nonumber
        & + & \left. \Delta_{\mu_2} 
                ( k_{2\mu_3} \Delta_{\mu_1} - \Delta_{\mu_3} k_{2\mu_1} )
                + \Delta \! \cdot \! k_2 ( g_{\mu_2 \mu_1} \Delta_{\mu_3}
                - g_{\mu_2 \mu_3} \Delta_{\mu_1} ) \right.\\ \nonumber
        & + & \left. \Delta_{\mu_3}
                ( k_{3\mu_1} \Delta_{\mu_2} - \Delta_{\mu_1} k_{3\mu_2} )
                + \Delta \! \cdot \! k_3 ( g_{\mu_3 \mu_2} \Delta_{\mu_1}
                - g_{\mu_3 \mu_1} \Delta_{\mu_2} ) \right] \, , \\
V_{\mu_1 \mu_2}^{a_1 a_2} (k_1,k_2) & = & \frac{1}{\alpha} \delta^{a_1 a_2} 
\left[ k_1^{\mu_1} \Delta^{\mu_2} \Delta \cdot k_1 + k_2^{\mu_2} \Delta^{\mu_1}
\Delta \cdot k_2 \right] \, , \\
W^{a_1 a_2} (k_1,k_2) & = & \delta^{a_1 a_2} \, \Delta \! \cdot \! k_1 \, 
		\Delta \! \cdot \! k_2 \, , \\
X^{a_1 a_2 a_3}_{\mu_3} (k_3) & = & -igf^{a_1 a_2 a_3} \, \Delta \! \cdot \! 
k_1 \, \Delta_{\mu_3} \, .
\end{eqnarray}
$T_{\mu_1 \mu_2}^{a_1 a_2} (k_1,k_2,M=2)$ is the Born level 
energy-momentum operator vertex and 
$V_{\mu_1 \mu_2}^{a_1 a_2} (k_1,k_2)$ vanishes when contracted 
with polarization vectors satisfying $\epsilon(k) \cdot k = 0$.

\section{Integrals}

In this appendix we briefly discuss some technicalities of the calculation.
For the case of $Q=0$ all integrals are expressible in terms of Eq. (\ref{I})
below.  The case of $Q \neq 0$ is more complicated and we discuss it now.
After partial fractioning the amplitudes we find everything is expressible 
in terms of the integrals
\begin{equation}
J(a,b,c,M)_{ \{ \;, \mu, \mu \nu \} } = \int 
	\frac{d^n k}{(2 \pi)^n} 
	\frac{ \{ 1, k_{\mu}, k_{\mu} k_{\nu} \} (\Delta \cdot k)^M}
	{(k^2)^a [ (k+p_1)^2 ]^b [ (k+p_2)^2 ]^c }\,.
\end{equation}
We write these integrals as linearly independent tensor structures 
multiplied by undetermined coefficients.
We then contract both sides with the various independent 
tensor structures and get a set of linear equations.  These we invert and 
solve for the previously undetermined coefficients.  At this point we have 
expressions for the tensor integrals above in terms of the scalar integral 
$J(a,b,c,M)$ and the tensor integrals with only two denominators, 
namely
\begin{equation}
\label{I}
I(a,b,M)_{ \{ \;, \mu, \mu \nu \} } 
	= \int \frac{d^n k}{(2 \pi)^n} 
	\frac{ \{ 1, k_{\mu}, k_{\mu} k_{\nu} \} (\Delta \cdot k)^M}
	{(k^2)^a [ (k+p)^2 ]^b }\,,
\end{equation}
which we again decompose as above.  The remaining integrals are expressible 
in terms of
\begin{eqnarray}
\label{Im}
I(a,b,M) & = & i S_n (-1)^{n/2} (p^2)^{n/2-a-b} (-1)^M 
	(\Delta \cdot p)^M \\ \nonumber
	& \times & \frac{\Gamma(a+b-n/2)}{\Gamma(a) \Gamma(b)}
	\frac{\Gamma(n/2-a+M) \Gamma(n/2-b)}{\Gamma(n+M-a-b)}\,,
\end{eqnarray}
and, for the $p_1^2 = p_2^2 = m^2$ case, the specific cases 
$J(1,1,1,0)$, $J(1,1,2,0)=J(1,2,1,0)$, $J(2,2,1,0)=J(2,1,2,0)$ 
and $J(2,1,1,0)$. 
In the above formulae $S_n = \pi^{n/2} / (2 \pi)^n$. 
The scalar $J$ integrals listed above can be found in the 
literature \cite{davy1}.  
We list the cases of interest here for completeness.  The 
integral $J(1,1,1,0)$ is finite in four dimensions for $m^2 \neq 0$:
\begin{equation}
J(1,1,1,0) = \frac{i}{16 \pi^2} \frac{1}{Q^2} x \Phi(x)
\end{equation}
where
\begin{eqnarray}
\Phi(x) & = & \frac{1}{\lambda(x)} \left\{ 2 \ln \left[ \frac{x-\lambda(x)}{2} 
\right] \ln \left[ \frac{2-x-\lambda(x)}{2} \right] \right. \\ \nonumber 
        & - & \left. 2 {\rm Li_2}  \left[ 
\frac{x-\lambda(x)}{2} \right]  - 2 {\rm Li_2}  \left[ 
\frac{2-x-\lambda(x)}{2} \right] + \frac{\pi^2}{3} \right\}
\end{eqnarray}
where ${\rm Li_2}$ is the dilogarithm function and
\begin{eqnarray}
\lambda(x) & = & x \sqrt{ 1 - 4/x } \\
x          & = & Q^2/m^2.
\end{eqnarray}
From the above and properties of the dilogarithm function one can work out 
that
\begin{equation}
Re [ x \Phi (x) ] \rightarrow \left\{ \begin{array}{ccc} \ln^2 ( x ) & 
	as & x \rightarrow \infty \\ \ln^2 ( -x ) & as & x \rightarrow -\infty
	\end{array} \right\}
\end{equation}
The other $J$ integrals have poles in $n-4$.  They are, dropping terms 
$O( n-4 )$,
\begin{eqnarray}
J(2,1,1,0) & = & i S_n \left[ \frac{2}{n-4} - \ln \left( \frac{-Q^2}{\mu^2}
\right) + 2 \ln \left( \frac{-m^2}{\mu^2} \right) + \gamma \right] 
\frac{1}{m^4} \\
J(1,2,1,0) & = & i S_n \left[ \frac{2}{n-4} + \ln \left( \frac{-Q^2}{\mu^2}
\right) + \gamma \right] \frac{1}{Q^2 m^2} \\
J(2,1,2,0) & = & i S_n \left[ \frac{2}{n-4} (Q^2+m^2) - Q^2 + 2 Q^2 
 \ln \left( \frac{-m^2}{\mu^2} \right) \right. \\ \nonumber 
        & + & \left. (m^2-Q^2) \ln \left( \frac{-Q^2}{\mu^2}
\right) + \gamma \right] \frac{1}{Q^2 m^6}
\end{eqnarray}
where $p_1^2=p_2^2=m^2$, $Q=p_1-p_2$, $\gamma$ is the 
Euler constant and the renormalization scale, 
which comes from the coupling constant implied on both sides of the 
formulae, is chosen to be $\mu^2$.
Note that the terms in $\ln ( \mu^2 )$ cancel in the final answers given 
in Section IV.

Whereas for the $p_1^2 = 0 = p_2^2$ case we need
\begin{eqnarray}
\label{Jm}
J(a,b,c,M) & = & i (-1)^{n/2} S_n (Q^2)^{n/2-a-b-c} (-\Delta \cdot p_2)^M 
\\ \nonumber
& &	\times
	\frac{ \Gamma(a+b+c-n/2) \Gamma(n/2-a-c) }{\Gamma(b) \Gamma(c) }
        \frac{ \Gamma(n/2+M-a-b) }{ \Gamma(n+M-a-b-c) } 
\\ \nonumber
& &     \times
	F_{2,1} \left( -M,\frac{n}{2}-a-c;1-\frac{n}{2}-M+a+b;
	\frac{\Delta \cdot p_1}{\Delta \cdot p_2} \right) \, ,
\end{eqnarray}
where $Q^2 = -2 p_1 \cdot p_2 \neq 0$, 
and $F_{2,1}(\alpha,\beta;\gamma;z)$ 
is the hypergeometric function of one variable.
Both scalar integrals Eq. (\ref{Im}) and Eq. (\ref{Jm}) 
can be derived by repeated 
differentiation of the $M=0$ integral or by the method of 
\cite{davy2}.  We have checked that both methods agree with each other 
and, for $M=2,4$ and 6, with the standard Passarino-Veltman 
reduction techniques \cite{pv}.  As stated in Section II, we use the 
condition $\Delta \cdot p_1 = \Delta \cdot p_2$.  Thus, the hypergeometric 
function in Eq. (\ref{Jm}) reduces to products of Gamma functions.

%
%

\end{document}